\documentclass[twocolumn,tighten,twocolappendix]{aastex63}
\pdfoutput=1 
\usepackage{amsmath,amstext}
\usepackage[T1]{fontenc}
\usepackage{ae,aecompl}
\usepackage[utf8]{inputenc}
\usepackage{apjfonts} 
\usepackage[figure,figure*]{hypcap}
\usepackage{enumitem}
\usepackage{bm}
\usepackage{graphicx}
\input{hyperlink-year-only-natbib-patch}

\usepackage{lineno}

\newcommand{\K}{\ensuremath{{\rm K}}}
\newcommand{\s}{\ensuremath{{\rm s}}}

\shorttitle{Joint Strong Lensing--Milky Way Satellite Analysis}
\shortauthors{Nadler et al.}

\begin{document}

\title{Dark Matter Constraints from a Unified Analysis of Strong Gravitational Lenses and Milky Way Satellite Galaxies}

\author[0000-0002-1182-3825]{Ethan O.~Nadler}
\affiliation{Department of Physics, Stanford University, 382 Via Pueblo Mall, Stanford, CA 94305, USA}
\affiliation{Kavli Institute for Particle Astrophysics \& Cosmology, P. O. Box 2450, Stanford University, Stanford, CA 94305, USA}
\author[0000-0002-3665-8141]{Simon Birrer}
\affiliation{Department of Physics, Stanford University, 382 Via Pueblo Mall, Stanford, CA 94305, USA}
\affiliation{Kavli Institute for Particle Astrophysics \& Cosmology, P. O. Box 2450, Stanford University, Stanford, CA 94305, USA}
\author[0000-0002-5116-7287]{Daniel Gilman}
\affiliation{Department of Astronomy and Astrophysics, University of Toronto, 50 St.\ George Street, Toronto, ON, M5S 3H4, Canada}
\author[0000-0003-2229-011X]{Risa H.~Wechsler}
\affiliation{Department of Physics, Stanford University, 382 Via Pueblo Mall, Stanford, CA 94305, USA}
\affiliation{Kavli Institute for Particle Astrophysics \& Cosmology, P. O. Box 2450, Stanford University, Stanford, CA 94305, USA}
\affiliation{SLAC National Accelerator Laboratory, Menlo Park, CA 94025, USA}
\author[0000-0003-0728-2533]{Xiaolong Du}
\affiliation{Carnegie Observatories, 813 Santa Barbara Street, Pasadena, CA 91101, USA}
\author[0000-0001-5501-6008]{Andrew Benson}
\affiliation{Carnegie Observatories, 813 Santa Barbara Street, Pasadena, CA 91101, USA}
\author[0000-0001-6809-2536]{Anna M.~Nierenberg}
\affiliation{Department of Physics, University of California Merced, 5200 North Lake Road, Merced, CA 95343, USA}
\author[0000-0002-8460-0390]{Tommaso Treu}
\affiliation{Department of Physics and Astronomy, University of California, Los Angeles, CA 90095-1547, USA}

\correspondingauthor{Ethan O.~Nadler}
\email{enadler@stanford.edu}

\begin{abstract}
Joint analyses of small-scale cosmological structure probes are relatively unexplored and promise to advance measurements of microphysical dark matter properties using heterogeneous data. Here, we present a multidimensional analysis of dark matter substructure using strong gravitational lenses and the Milky Way (MW) satellite galaxy population, accounting for degeneracies in model predictions and using covariances in the constraining power of these individual probes for the first time. We simultaneously infer the projected subhalo number density and the half-mode mass describing the suppression of the subhalo mass function in thermal relic warm dark matter (WDM), $M_{\mathrm{hm}}$, using the semianalytic model \texttt{Galacticus} to connect the subhalo population inferred from MW satellite observations to the strong lensing host halo mass and redshift regime. Combining MW satellite and strong lensing posteriors in this parameter space yields $M_{\mathrm{hm}}<10^{7.0}\ M_{\mathrm{\odot}}$ (WDM particle mass $m_{\mathrm{WDM}}>9.7\ \mathrm{keV}$) at $95\%$ confidence and disfavors $M_{\mathrm{hm}}=10^{7.4}\ M_{\mathrm{\odot}}$ ($m_{\mathrm{WDM}}=7.4\ \mathrm{keV}$) with a 20:1 marginal likelihood ratio, improving limits on $m_{\mathrm{WDM}}$ set by the two methods independently by $\sim 30\%$. These results are marginalized over the line-of-sight contribution to the strong lensing signal, the mass of the MW host halo, and the efficiency of subhalo disruption due to baryons and are robust to differences in the disruption efficiency between the MW and strong lensing regimes at the $\sim 10\%$ level. This work paves the way for unified analyses of next-generation small-scale structure measurements covering a wide range of scales and redshifts.
\end{abstract}

\keywords{\href{http://astrothesaurus.org/uat/353}{Dark matter (353)}; \href{http://astrothesaurus.org/uat/261}{Strong gravitational lensing (261)}; \href{http://astrothesaurus.org/uat/1049}{Milky Way dark matter halo (1049)};
\href{http://astrothesaurus.org/uat/574}{Galaxy abundances (574)}}


\section{Introduction}
\label{Introduction}

The $\Lambda$CDM cosmological paradigm assumes a cold, collisionless dark matter (CDM) particle and therefore predicts a plethora of dark matter structure and substructure on extremely small cosmic scales (e.g., \citealt{Green0309621,Diemand0501589,Wang191109720}). It is often argued that small-scale structure measurements represent an outstanding test to this prediction (e.g., see \citealt{Bullock170704256} for a review); yet, our understanding of the distribution of dark matter structure on nonlinear scales is rapidly progressing. Recent analyses of Milky Way (MW) satellite galaxies using data over nearly the full sky---including the population of ultrafaint dwarf galaxies discovered by deep photometric surveys over the last decade---have only recently been performed (e.g., \citealt{Drlica-Wagner191203302,Nadler191203303,Nadler200800022}). Meanwhile, measurements of stellar streams from the Gaia mission are beginning to reach the requisite precision to infer the signatures of perturbations from nearby low-mass subhalos \citep{Banik191102662,Bonaca181103631}. On extragalactic scales, the number of compact-source strong gravitational lenses available for substructure analyses has drastically increased in recent years (e.g., \citealt{Nierenberg14021496,Nierenberg170105188,Nierenberg190806344}), and modeling efforts have advanced in step (e.g., \citealt{Gilman190111031,Gilman190806983,Hsueh190504182}). Analyses of resolved distortion in extended strong lensing observations from adaptive optics and space-based imaging have also rapidly progressed (e.g., \citealt{Hezaveh160101388,Birrer17020009,Vegetti180101505}).

All of the recent small-scale structure measurements outlined above are consistent with the CDM paradigm and have therefore been used to constrain microphysical properties of dark matter that would reduce its small-scale clustering \citep{Banik191102662,Gilman190806983,Nadler200800022}. Although analyses of different probes reach consistent dark matter constraints, to date they have been performed independently and with different modeling assumptions to address heterogeneous astrophysical systematics. Crucially, if evidence for a departure from the CDM paradigm arises, it must be confirmed across different redshifts and physical scales. It is therefore critical to jointly model and analyze small-scale structure probes. This effort will be particularly important to maximize small-scale structure measurements from next-generation surveys including the Rubin Observatory Legacy Survey of Space and Time (LSST), which will enable the discovery of vastly more strong gravitational lenses (e.g., \citealt{Collett150702657}) and revolutionize the search for dwarf galaxies and measurements of stellar streams in the local universe (e.g., \citealt{Drlica-Wagner190201055}).

Jointly modeling the low-mass halo and subhalo populations relevant for various small-scale structure measurements requires precise theoretical predictions for the abundance and structure of these small systems---which probe highly nonlinear cosmological modes---as a function of redshift and environment. Even cosmological parameters play an important role given the precision of current data; for example, varying the running of the spectral index within Planck uncertainties significantly affects predictions for subhalo abundances \citep{Stafford200403872}, while other cosmological parameters including $\Omega_{m}$ and $\sigma_8$ have subleading effects that may become important to incorporate in models of next-generation small-scale structure data \citep{Dooley14036828}. Moreover, a variety of other theoretical and numerical uncertainties must be marginalized over in joint likelihood analyses to robustly claim evidence for non-CDM physics. For example, specific systematics of interest for modeling the MW satellite galaxy population include the faint end of the galaxy--halo connection, the total mass of the MW halo, and the mass and accretion time of the Large Magellanic Cloud (LMC; \citealt{Newton201108865,Nadler200800022}). Meanwhile, the orbits of dark matter subhalos in the inner regions of the MW halo must be predicted precisely in a statistical sense while accurately modeling the effects of specific baryonic structures to infer dark matter properties from stellar stream measurements. For strong lensing, the mass--concentration relation in both CDM and alternative dark matter models is a key uncertainty that must be accounted for \citep{Hezaveh160101388,Gilman190806983,Gilman190902573,Minor201110629}, along with the host halo properties and selection functions of strong lenses, all while accurately modeling the differential signal contributed by substructure and small halos along the line of sight \citep{Despali171005029,Gilman190111031}.

Here, we perform a joint analysis of small-scale dark matter measurements by combining the results of recent strong gravitational lensing and MW satellite inferences in a multidimensional parameter space to break modeling degeneracies.\footnote{A joint small-scale structure analysis by \cite{Enzi201013802} appeared during the preparation of this manuscript. We comment on the differences between the methodology and results of our study and this work in Section \ref{sec:study}.} In particular, we combine these results in a parameter space that includes the mass scale describing the suppression of the subhalo mass function for thermal relic warm dark matter (WDM) and the amplitude of the projected subhalo mass function at the strong lensing host halo mass and redshift scale. In particular, we combine the constraints on these quantities derived from (i) the magnification and flux ratio data from quadruply imaged strong gravitational lenses presented in \cite{Nierenberg14021496,Nierenberg170105188,Nierenberg190806344} and modeled in \cite{Gilman190806983}, and (ii) the abundance and properties of MW satellite galaxies over $\sim 75\%$ of the sky presented in \cite{Drlica-Wagner191203302} and modeled in \cite{Nadler191203303,Nadler200800022}. We employ the semianalytic model \texttt{Galacticus} \citep{Benson10081786,Pullen14078189} to translate the subhalo population inferred from MW satellite measurements to the strong lensing host halo mass and redshift scale by calibrating its predictions using cosmological zoom-in simulations of MW-like halos \citep{Mao150302637}. Thus, our work lays the foundation for joint semianalytic models of small-scale structure that are benchmarked by high-resolution simulations at each halo mass and redshift scale.

Our joint analysis breaks degeneracies among the amplitude of the projected subhalo mass functions inferred from MW satellite and strong lensing observations, thereby improving limits on deviations from the CDM paradigm that have been derived independently from these data. Specifically, we show that our combined analysis improves the lower limit on the WDM particle mass by \textcolor{black}{$\sim 30\%$}. The framework we develop for combining MW satellite and strong lensing data is particularly important because strong lensing is potentially sensitive to the presence of halos with masses below the threshold for galaxy formation, a mass scale that dwarf galaxy observations constrain. We therefore quantify the observational and theoretical advances necessary to robustly infer the presence of such dark halos, showing that this outcome is within the reach of next-generation small-scale structure measurements.

This paper is organized as follows. In Section \ref{sec:overview}, we describe the analytic model of dark matter substructure that underlies our joint analysis. We then describe the MW satellite data and model that enters our analysis in Section \ref{sec:mw} and the strong lensing data and model in Section \ref{sec:lens}. We combine these analyses in Section \ref{sec:bridge}, present our results in Section \ref{sec:results}, discuss key systematics and compare them to previous work in Section \ref{sec:discussion}, and conclude in Section \ref{sec:conclusion}. Throughout, we adopt the following cosmological parameters, following both \cite{Gilman190806983} and \cite{Nadler200800022}: $h = 0.7$, $\Omega_{\rm m} = 0.286$, $\Omega_{\Lambda} = 0.714$, $\sigma_8 = 0.82$, and $n_s=0.96$ \citep{Hinshaw_2013}.


\section{Dark Matter Substructure Model}
\label{sec:overview}

We begin by describing the analytic model of dark matter substructure used to connect the subhalo populations probed by MW satellite and strong lensing observations. In particular, we describe our model for the projected subhalo mass function (SHMF; Section \ref{sec:projected_shmf}), its dependence on host halo mass and redshift (Section \ref{sec:cdm_dependence}), and the efficiency of subhalo disruption due to baryons (Section \ref{sec:q_description}). We then describe our model for the impact of WDM physics on subhalo abundances (Section \ref{sec:wdm_mhm}) and concentrations (Section \ref{sec:wdm_mc}).

\subsection{Projected subhalo mass function}
\label{sec:projected_shmf}

Strong lensing and MW satellites probe low-mass subhalos within host halos at different mass and redshift scales. Specifically, strong lensing probes both the projected dark matter substructure in the lens system and small-scale structure along the line of sight, while MW satellites probe the three-dimensional distribution of subhalos traced by luminous satellite galaxies within the MW. Because current strong lensing analyses are not highly sensitive to the radial distribution of subhalos within the host halo of the lens, we focus on the statistics of projected subhalo populations in this paper, although we will describe how observations of the radial distribution of MW satellites break model degeneracies.

To simultaneously predict the subhalo populations relevant for MW satellite and strong lensing studies, we construct an analytic model for projected subhalo abundances that depends on the host halo mass, $M_{\mathrm{host}}$, and redshift, $z_{\mathrm{host}}$. In particular, we express the projected SHMF---i.e., the differential number of subhalos within a host halo, in projection---by generalizing the form in \cite{Gilman190806983},
\begin{equation}
    \frac{\mathrm{d}^2 N_{\mathrm{sub}}}{\mathrm{d}M \mathrm{d}A} \equiv \frac{\Sigma_{\mathrm{sub}}(M_{\mathrm{host}},z_{\mathrm{host}})}{M_0}\left(\frac{M}{M_0}\right)^{\alpha}  \mathcal{F}_{\mathrm{CDM}}(M_{\mathrm{host}},z_{\mathrm{host}}),
\label{eq:shmf}
\end{equation}
where $M$ denotes subhalo mass, $A$ denotes the unit area, $M_0=10^8\ M_{\mathrm{\odot}}$ and $\alpha$ is the power-law slope of the SHMF. In Equation \ref{eq:shmf}, $\Sigma_{\mathrm{sub}}(M_{\mathrm{host}},z_{\mathrm{host}})$ is the projected number density of subhalos within the virial radius of a host halo of mass $M_{\mathrm{host}}$ at redshift $z_{\mathrm{host}}$, including the effects of baryonic physics, and $\mathcal{F}_{\mathrm{CDM}}(M_{\mathrm{host}},z_{\mathrm{host}})$ captures the dependence of the projected SHMF on $M_{\mathrm{host}}$ and $z_{\mathrm{host}}$ in CDM only (i.e., modulo the effects of baryonic physics) as described in Section \ref{sec:cdm_dependence}. We discuss the impact of halo boundary definitions in Section \ref{sec:bridge}.

Our model of the MW satellite population is based on the \cite{Nadler200800022} analysis, which defines subhalo mass using the peak Bryan--Norman virial mass $M_{\mathrm{peak}}$ (see Appendix \ref{sec:sim_details} for details). Meanwhile, our strong lensing constraints are based on the \cite{Gilman190806983} analysis, which uses $M_{200}$ values relative to the critical density at $z=0$, with subhalo masses evaluated at infall to compute the WDM SHMF and mass--concentration relation. Here, we simply interpret the peak virial mass values from our MW satellite analysis as $M_{200}$ values at infall. The peak virial masses of subhalos relevant for our work are on average a factor of $\sim 2$ larger than their $M_{200}$ values at infall in the cosmological zoom-in simulations our MW satellite analysis is based on, largely due to pre-infall tidal stripping (e.g., \citealt{Behroozi13102239,Wetzel150101972}). Thus, converting $M_{\mathrm{peak}}$ to $M_{200}$ would further strengthen our joint WDM constraints; however, because changing this mass definition would nontrivially affect the abundance-matching model used in our satellite analysis, we leave a detailed investigation of this point to future work that combines satellite and lensing inferences at the likelihood level.

\subsection{CDM host mass and redshift dependence}
\label{sec:cdm_dependence}

We model the dependence of the projected SHMF on host halo mass and redshift with the functions $\Sigma_{\mathrm{sub}}(M_{\mathrm{host}},z_{\mathrm{host}})$ and $\mathcal{F}_{\mathrm{CDM}}(M_{\mathrm{host}},z_{\mathrm{host}})$. Both of these terms play a key role in our joint analysis because they allow us to relate the subhalo populations corresponding to low-redshift, group-mass strong lens host halos ($M_{\mathrm{lens}}\sim 10^{13}\ M_{\mathrm{\odot}}$, $z_{\mathrm{lens}}\sim 0.5$) to the regime of the MW halo today ($M_{\mathrm{MW}}\sim 10^{12}\ M_{\mathrm{\odot}}$, $z_{\mathrm{MW}}=0$).

$\mathcal{F}_{\mathrm{CDM}}(M_{\mathrm{host}},z_{\mathrm{host}})$ captures the dependence of the SHMF on host halo mass and redshift in CDM only (i.e., without baryons), including the effects of tidal disruption by the dark matter host halo. This scaling depends on both the statistics of subhalo populations at infall, which can be predicted reasonably precisely using extensions of the Press--Schechter formalism \citep{EPS}, and on the dynamical evolution of subhalos after infall into a host. Detailed semianalytic models calibrated to $N$-body simulations are necessary to model this evolution; we therefore follow \cite{Gilman190806983} in using the \texttt{Galacticus} model \citep{Benson10081786,Pullen14078189,Yang200310646}, which predicts
\begin{align}
    \log\mathcal{F}_{\mathrm{CDM}}(M_{\mathrm{host}},z_{\mathrm{host}}) =~ &k_1 \log\left(\frac{M_{\mathrm{host}}}{10^{13}\ M_{\mathrm{\odot}}}\right)&\nonumber\\ &+ k_2 \log(z_{\mathrm{host}}+0.5),&
    \label{eq:galacticus}
\end{align}
where $k_1=0.88$ and $k_2=1.7$. 

Section \ref{sec:q_description} describes our model for subhalo disruption due to baryons, which captures the leading-order corrections to $\mathcal{F}_{\mathrm{CDM}}(M_{\mathrm{host}},z_{\mathrm{host}})$. We do not model the impact of additional host halo, central galaxy, and environmental variables on the projected SHMF, noting that this is an important area for future work that ongoing observational efforts like the Satellites Around Galactic Analogs (SAGA) survey are informing at the MW-mass scale \citep{Geha170506743,Mao200812783}. However, we emphasize that the \cite{Nadler200800022} MW satellite analysis our work is based on self-consistently uses simulations that are consistent with key secondary MW halo properties, including concentration, the existence of a realistic LMC analog system, and a formation history constrained by Gaia observations. Meanwhile, the \cite{Gilman190806983} lensing analysis is not sensitive to host-to-host variation in $\Sigma_{\mathrm{sub}}$ beyond that modeled by $\mathcal{F}_{\mathrm{CDM}}(M_{\mathrm{host}},z_{\mathrm{host}})$ given the current number of strong lenses studied and the information available per lens.

\subsection{Subhalo disruption efficiency due to baryons}
\label{sec:q_description}

We model $\mathcal{\Sigma_{\mathrm{sub}}}(M_{\mathrm{host}},z_{\mathrm{host}})$ with explicit host halo mass and redshift dependence to capture the impact of baryonic physics on the projected SHMF. This extra dependence relative to the CDM scaling is not captured by the $\mathcal{F}_{\mathrm{CDM}}(M_{\mathrm{host}},z_{\mathrm{host}})$ term predicted by \texttt{Galacticus}, although baryonic effects can be modeled in future \texttt{Galacticus} implementations. Although $\Sigma_{\mathrm{sub}}$ is not modeled with explicit host mass and redshift dependence in \cite{Gilman190806983}, we include this dependence here because the subhalo populations probed by strong lensing and MW satellites are subject to baryonic effects that potentially impact the two regimes differently. Of these effects, the most important is tidal disruption due to the central galaxy, which suppresses the SHMF at the $\sim 50\%$ level (e.g., \citealt{Despali160806938,Garrison-Kimmel170103792,Graus171011148,Kelley181112413,Richings181112437,Webb200606695}).\footnote{Supernova feedback within sufficiently massive subhalos can also reduce their inner densities (e.g., \citealt{Governato12020554,Pontzen11060499,Read180806634}) and accelerate disruption, but hydrodynamic simulations suggest that this process has a subleading effect on the SHMF compared to disruption by the central galaxy.} Tidal disruption due to the central galaxy most strongly suppresses the abundance of subhalos in the inner regions of the host halo or (more precisely) subhalos that accrete early and have close pericentric passages \citep{Garrison-Kimmel170103792,Nadler171204467}.

The projected SHMF is largely driven by the plethora of subhalos in the outer regions of the host halo, which mitigates the impact of uncertainties in the strength and radial dependence of these baryonic effects on our probe combination. Nonetheless, our joint analysis is sensitive to both the amplitude of and differences in the efficiency of subhalo disruption due to baryonic physics as a function of host halo mass and redshift. We measure $\Sigma_{\mathrm{sub}}(M_{\mathrm{host}},z_{\mathrm{host}})$ in units of its value for strong lens host halos, and we define the differential subhalo disruption efficiency due to baryons as
\begin{equation}
    q \equiv \frac{\Sigma_{\mathrm{sub}}(M_{\mathrm{MW}},z_{\mathrm{MW}})}{\Sigma_{\mathrm{sub}}(M_{\mathrm{lens}},z_{\mathrm{lens}})} \equiv \frac{\Sigma_{\mathrm{sub,MW}}}{\Sigma_{\mathrm{sub}}},\label{eq:q}
\end{equation}
where $\Sigma_{\mathrm{sub}}$ hereafter denotes the projected subhalo number density for strong lens host halos, following \cite{Gilman190806983}, and $\Sigma_{\mathrm{sub,MW}}$ denotes the same quantity for the present-day MW system. In Equation \ref{eq:q}, $q$ represents the efficiency of subhalo disruption due to baryonic physics in the MW at $z=0$ in units of the efficiency of subhalo disruption due to baryonic physics in the group-mass halos and at the redshifts probed by strong lensing. Note that larger (smaller) values of $q$ represent less efficient (more efficient) subhalo disruption in the MW relative to strong lenses and that differences in the radial dependence of subhalo disruption at these scales (which we do not model) do not affect our joint analysis of projected SHMFs.

Motivated by the results of hydrodynamic simulations, we assume that subhalo disruption due to baryonic physics results in a mass-independent rescaling of the MW and strong lens projected SHMFs. This allows us to respectively express the projected SHMFs probed by strong lensing and MW satellite observations as
\begin{flalign}
&\left(\frac{\mathrm{d}^2 N_{\mathrm{CDM}}}{\mathrm{d}M \mathrm{d}A}\right)_{\mathrm{lensing}} = \frac{\Sigma_{\mathrm{sub}}}{M_0}\left(\frac{M}{M_0}\right)^{\alpha} \mathcal{F}_{\mathrm{CDM}}(M_{\mathrm{lens}},z_{\mathrm{lens}}),\label{eq:shmf_lensing}&
\end{flalign}
\begin{flalign}
\left(\frac{\mathrm{d}^2 N_{\mathrm{CDM}}}{\mathrm{d}M \mathrm{d}A}\right)_{\mathrm{MW}} &= \frac{\Sigma_{\mathrm{sub,MW}}}{M_0}\left(\frac{M}{M_0}\right)^{\alpha}  \mathcal{F}_{\mathrm{CDM}}(M_{\mathrm{MW}},z_{\mathrm{MW}})&\nonumber\\ &= \frac{q\Sigma_{\mathrm{sub}}}{M_0}\left(\frac{M}{M_0}\right)^{\alpha}  \mathcal{F}_{\mathrm{CDM}}(M_{\mathrm{MW}},z_{\mathrm{MW}}).&
\label{eq:shmf_mw}
\end{flalign}

As noted above, strong lenses typically have halo masses of $M_{\mathrm{lens}}\approx 10^{13}\ M_{\mathrm{\odot}}$ and redshifts of $z_{\mathrm{lens}}\approx 0.5$, and host massive elliptical galaxies \citep{Gavazzi0701589,Auger10072880,Gilman190806983}. In contrast, the MW has a halo mass of $M_{\mathrm{MW}}\sim 10^{12}\ M_{\mathrm{\odot}}$ (e.g., \citealt{Callingham180810456,Cautun191104557}) at $z_{\mathrm{MW}}=0$, and is largely typical for a spiral galaxy of its stellar mass, although it has a relatively quiescent formation history (e.g., \citealt{Boardman200902576,Evans200504969}). Subhalo disruption due to the central galaxy in hydrodynamic simulations of MW-mass systems reduces the amplitude of the SHMF by $\sim 50\%$ relative to corresponding dark-matter-only simulations. This effect is roughly mass independent and is not a strong function of redshift at late times. Although hydrodynamic simulations of group-mass systems yield similar levels of SHMF suppression \citep{Fiacconi160203526,Graus171011148,Richings200514495}, this regime is less well studied. We therefore adopt $q=1$ in our fiducial analysis---i.e., equally efficient subhalo disruption due to baryons in the MW and strong lens host halos---and we also test values within a range of $q\in [0.5,2]$ when translating the projected SHMF amplitude inferred from MW satellites to the strong lens host halo regime. We emphasize that, in detail, subhalo disruption is expected to depend on the mass and formation history of the central galaxy along with host halo mass and redshift, and its efficiency will therefore differ among strong lenses. Although our phenomenological model for differences in subhalo disruption due to baryonic physics is very simple, we will demonstrate that the corresponding uncertainties do not significantly impact our joint dark matter constraints.

\subsection{Warm dark matter subhalo mass function}
\label{sec:wdm_mhm}

The half-mode mass, $M_{\mathrm{hm}}$, represents a characteristic mass scale describing the suppression of the linear matter power spectrum due to non-CDM physics; in particular, it corresponds (in linear theory) to the wavenumber at which the ratio of the linear matter power spectrum drops to $25\%$ of that in CDM (e.g., \citealt{Nadler190410000}). In the case of thermal relic WDM, free streaming suppresses the power spectrum on small scales, leading to a turnover in the halo and subhalo mass functions below $M_{\rm{hm}}$, which in turn depends on the WDM particle mass, $m_{\mathrm{WDM}}$ (e.g., \citealt{Schneider11120330}). MW satellites constrain this suppression by tracing the abundance of low-mass halos, while the subhalos surrounding the main deflector in strong lenses affect image flux ratios.

The WDM SHMF can be expressed as
\begin{equation}
\frac{\mathrm{d}N_{\rm{WDM}}}{\mathrm{d}M} \equiv f_{\mathrm{WDM}}\left(M, M_{\rm{hm}}\right) \frac{dN_{\rm{CDM}}}{dM},\label{eq:wdm_shmf}
\end{equation}
where $\mathrm{d}N_{\rm{WDM}}/\mathrm{d}M$ ($\mathrm{d}N_{\rm{CDM}}/\mathrm{d}M$) is the WDM (CDM) SHMF, and $f_{\mathrm{WDM}}$ is a multiplicative suppression factor that depends on subhalo mass $M$ and the WDM particle mass $m_{\mathrm{WDM}}$ via $M_{\mathrm{hm}}$. We follow both \cite{Gilman190806983} and \cite{Nadler200800022} by assuming that this SHMF suppression does not alter the (normalized) radial distribution of subhalos, consistent with the findings of WDM simulations (e.g., \citealt{Lovell13081399,Bose++16}). Thus, the same multiplicative factor $f_{\mathrm{WDM}}\left(M, M_{\rm{hm}}\right)$ dictates the suppression of the projected SHMFs in our model, i.e.,
\begin{flalign}
    &\left(\frac{\mathrm{d}^2 N_{\mathrm{WDM}}}{\mathrm{d}M \mathrm{d}A}\right)_{\mathrm{lensing}} = f_{\mathrm{WDM}}(M,M_{\mathrm{hm}})\left(\frac{\mathrm{d}^2 N_{\mathrm{CDM}}}{\mathrm{d}M \mathrm{d}A}\right)_{\mathrm{lensing}},&\\
     &\left(\frac{\mathrm{d}^2 N_{\mathrm{WDM}}}{\mathrm{d}M \mathrm{d}A}\right)_{\mathrm{MW}} = f_{\mathrm{WDM}}(M,M_{\mathrm{hm}})\left(\frac{\mathrm{d}^2 N_{\mathrm{CDM}}}{\mathrm{d}M \mathrm{d}A}\right)_{\mathrm{MW}}.&
\end{flalign}

For concreteness and to allow for an apples-to-apples comparison between lensing and MW satellite analyses, we focus on the case of thermal relic WDM, for which the SHMF can be expressed as \citep{Lovell200301125}
\begin{equation}
f_{\mathrm{WDM}}(M,M_{\mathrm{hm}}) = \left[1+\left(\frac{\alpha M_{\mathrm{hm}}(m_{\mathrm{WDM}})}{M}\right)^{\beta}\right]^{\gamma},\label{eq:wdm_suppression}
\end{equation}
where $M_{\mathrm{hm}}$ is related to $m_{\mathrm{WDM}}$ in our fiducial cosmology via \citep{Nadler200800022}
\begin{equation}
M_{\mathrm{hm}}(m_{\mathrm{WDM}}) = 5 \times 10^8 \left(\frac{m_{\mathrm{WDM}}}{3\  \mathrm{keV}}\right)^{-10/3} M_{\mathrm{\odot}}.
\label{eq:mhm_mwdm}
\end{equation}
In Equation \ref{eq:wdm_suppression}, $\alpha$, $\beta$, and $\gamma$ are free parameters fit to simulation results. The analysis in \cite{Nadler200800022} uses the SHMF from \cite{Lovell13081399}, which corresponds to $\alpha=2.7$, $\beta=1.0$, and $\gamma=-0.99$, while \cite{Gilman190806983} adopt an alternative fit to the SHMF from \cite{Lovell13081399}, corresponding to $\alpha=1$, $\beta=1$, and $\gamma=-1.3$. As described in Section \ref{sec:mw}, we rerun the MW satellite analysis with the \cite{Gilman190806983} choice of WDM SHMF suppression in order to self-consistently combine the posterior distributions from these analyses according to the procedure in Section \ref{sec:probe_combination_stats}.

\subsection{Warm dark matter mass--concentration relation}
\label{sec:wdm_mc}

The delay in the collapse of small-scale density perturbations in WDM suppresses the central densities of halos with masses near $M_{\rm{hm}}$, altering the mass--concentration relation for both field and subhalos. Because flux ratios in strong lenses are highly sensitive to the central densities of subhalos, the altered mass--concentration relation provides crucial information relevant for forward-modeling strong lensing signals \citep{Gilman190902573}. We implement the WDM mass--concentration relation in a similar manner to the suppression of the SHMF \citep{Gilman190806983},
\begin{equation}
c_{\rm{WDM}}\left(M\right) \equiv f_{\mathrm{WDM}}^{\prime}\left(M, M_{\rm{hm}}\right) c_{\rm{CDM}}\left(M\right),
\end{equation}
where $c_{\rm{WDM}}\left(M\right)$ ($c_{\rm{CDM}}\left(M\right)$) is the WDM (CDM) mass--concentration relation, and $f'_{\mathrm{WDM}}$ is a concentration suppression factor analogous to $f_{\mathrm{WDM}}$. In particular, we follow \cite{Gilman190806983} by using $c_{\mathrm{CDM}}(M)$ from \cite{Diemer180907326} with $0.1\ \mathrm{dex}$ scatter \citep{Dutton14027073} and $c_{\mathrm{WDM}}(M)$ from \cite{Bose++16}, 
\begin{equation}
f^{\prime}_{\mathrm{WDM}}\left(M,M_{\rm{hm}}\right) = \left(1+z\right) ^{\beta\left(z\right)} \left(1+60\frac{M_{\rm{hm}}}{M}\right)^{-0.17},
\end{equation}
where $\beta(z) = 0.026z - 0.04$.

Halo concentrations are affected over an order of magnitude in mass above the turnover in the mass function set by $M_{\rm{hm}}$. Thus, the mass--concentration relation must be accounted for to self-consistently constrain WDM-like models using strong lensing data. Meanwhile, MW satellite abundances are relatively insensitive to the mass--concentration relation because subhalo disruption is mainly determined by subhalos' orbital properties. Moving beyond abundances, the internal dynamics of relatively bright MW satellite galaxies are often subject to baryonic effects that make it difficult to robustly infer halo concentration (e.g., \citealt{Read180806634}). Meanwhile, it is difficult to obtain precise dynamical measurements given the limited number of spectroscopically confirmed stars associated with the faintest MW satellites; however, future spectroscopic measurements of these galaxies may reach the precision necessary to provide complementary constraints \citep{Simon190304743}.


\section{Milky Way Substructure Modeling}
\label{sec:mw}

We now review the key components of the \cite{Nadler191203303,Nadler200800022} MW satellite analyses our study uses. These analyses, which respectively constrain the galaxy--halo connection in CDM and non-CDM scenarios, are based on a forward model of the MW satellite population that combines high-resolution simulations of halos selected to resemble the MW combined with an empirical model for the galaxy--halo connection. These studies account for observational selection functions to fit the MW satellite population in a statistical framework and infer the underlying SHMF, which in turn constrains dark matter physics. For brevity, we mainly describe the \cite{Nadler200800022} WDM analysis, and we refer the reader to specific sections of \cite{Nadler191203303} for further methodological details throughout the following subsections.

\subsection{Milky Way Satellite Data}

\cite{Nadler200800022} analyze the kinematically confirmed and candidate MW satellites from \cite{Drlica-Wagner191203302}, which were identified using Dark Energy Survey (DES) and Pan-STARRS1 (PS1) data. In particular, \cite{Nadler200800022} analyze 34 satellite galaxies with stellar masses from $\sim 10^2\ M_{\mathrm{\odot}}$ to $10^7\ M_{\mathrm{\odot}}$. Together, the DES and PS1 datasets cover more than $\sim 75\%$ of the high-Galactic-latitude sky and provide exquisite sensitivity near the LMC due to deep DES photometry in that region. Thus, \cite{Nadler200800022} incorporate both inhomogeneity and incompleteness in the observed MW satellite population through by using the observational selection functions from \cite{Drlica-Wagner191203302}. Unlike other semiempirical models of the MW satellite population (e.g., \citealt{Jethwa161207834,Kim1812121,NewtonMNRAS,Newton201108865}), \cite{Nadler200800022} account for the effect of the LMC system on the observed MW satellite population, which is essential to fit the full dataset.

\subsection{Milky Way Satellite Model}

\subsubsection{Milky Way Zoom-in Simulations}

The MW satellite model used in \cite{Nadler200800022} is based on high-resolution dark-matter-only zoom-in simulations selected from the suite of 45 MW-mass hosts presented in \cite{Mao150302637}; technical details on these simulations, which resolve subhalos with virial masses as small as $\sim 10^7\ M_{\mathrm{\odot}}$ at $z=0$, are provided in Appendix \ref{sec:sim_details}. In particular, \cite{Nadler200800022} use the two most ``MW-like'' host halos in this simulation suite to model the MW satellite population. These hosts have mass and concentration values consistent with recent inferences based on Gaia data \citep{Callingham180810456,Cautun191104557}. In addition, they have early major mergers that resemble the Gaia-Enceladus event as well as nearby, recently accreted LMC analogs that match the satellite population and kinematics of the real LMC system (see \citealt{Nadler191203303} Section 7.2).

\subsubsection{Galaxy--Halo Connection Model}

To infer the present-day abundance of subhalos in the MW, \cite{Nadler200800022} combine the simulations described above with an empirical model of the galaxy--halo connection (introduced in \citealt{Nadler180905542,Nadler191203303}), which populates subhalos with satellite galaxies in a parametric fashion. By combining these predictions with observational selection functions derived from satellite searches in DES and PS1 data \citep{Drlica-Wagner191203302}, the model is compared to observations assuming that satellites in each survey footprint populate the parameter space of surface brightness and heliocentric distance according to a Poisson process (see \citealt{Nadler191203303} Section 6). By marginalizing over the underlying Poisson rate in the calculation of the likelihood for each surface brightness bin, the galaxy--halo connection and dark matter model parameters are fit to data in a Markov Chain Monte Carlo (MCMC) framework.

The majority of the parameters in the \cite{Nadler200800022} WDM analysis govern the relationship between satellite galaxies and the subhalos they inhabit. For example, these include the slope and scatter of the abundance-matching relation between galaxy luminosity and peak halo maximum circular velocity; the amplitude, scatter, and power-law slope of the relation between galaxy size and halo size; and parameters governing the fraction of low-mass dark matter halos that host observable galaxies. These parameters are not directly relevant for our strong lensing joint analysis because lensing is sensitive to the integrated amount of matter in the lens galaxy and along the line of sight, which is dominated by dark matter. However, they are crucial for robustly modeling the MW satellite population and are marginalized over in our probe combination.

Here, we highlight the aspects of the \cite{Nadler200800022} model that are most relevant for our joint analysis:
\begin{enumerate}[wide, label=({\emph{\roman*}}), itemsep=0pt]
    \item \emph{Minimum halo mass} ($\mathcal{M}_{\mathrm{min}}$): The \cite{Nadler200800022} satellite analysis is consistent with CDM predictions down to a characteristic halo mass scale referred to as the minimum halo mass ($\mathcal{M}_{\mathrm{min}}$). The minimum halo mass is defined as the peak virial mass of the smallest surviving subhalo inferred to host observed MW satellite galaxies and therefore represents the lowest mass down to which the SHMF is directly constrained by current MW satellite observations. $\mathcal{M}_{\mathrm{min}}$ is jointly inferred along with the fraction of halos that host observable galaxies, which is consistent with $100\%$ down to $\mathcal{M}_{\mathrm{min}}$. The upper limit on $\mathcal{M}_{\mathrm{min}}$ is calculated by marginalizing over the full posterior distribution, which yields~$\mathcal{M}_{\mathrm{min}}<3.2\times 10^8\ M_{\mathrm{\odot}}$ at $95\%$ confidence in the \citep{Nadler191203303} CDM fit (see \citealt{Nadler191203303} Sections 4.4, 7.4, and 7.5).
    
    \item \emph{Baryonic disruption efficiency} ($\mathcal{B}$): The efficiency of subhalo disruption due to the Galactic disk. Disruption probabilities due to baryonic physics for the subhalos in the dark-matter-only simulations described above are predicted using the \cite{Nadler171204467} subhalo disruption model, which is calibrated to hydrodynamic simulations from the Feedback in Realistic Environments (FIRE) project \citep{Garrison-Kimmel170103792}. We note that several subsequent DM-only plus disk \citep{{Kelley181112413}} and hydrodynamic simulations \citep{Richings181112437,Samuel190411508} of MW-mass halos report comparable amounts of subhalo disruption relative to the FIRE simulations used to calibrate the \cite{Nadler171204467} disruption model.
    
    To account for uncertainties resulting from the limited statistics of these training simulations, \cite{Nadler200800022} parameterize the efficiency of subhalo disruption by assigning the following disruption probability to each subhalo in the MW-like zoom-in simulations:
    \begin{equation}
    p_{\mathrm{disrupt}} \equiv (p_{\mathrm{disrupt,0}})^{1/\mathcal{B}},
    \end{equation}
    where $p_{\mathrm{disrupt,0}}$ is the fiducial disruption probability predicted by the \cite{Nadler171204467} model, which is a function of the orbital properties (including pericentric distance and accretion time) and internal properties (including mass and maximum circular velocity at accretion) of subhalos. Adopting a lognormal prior on $\mathcal{B}$ centered on the hydrodynamic training simulations (i.e., $\mathcal{B}=1$), the WDM MW satellite analysis in \cite{Nadler200800022} yields disruption efficiencies that are consistent with hydrodynamic simulations and rule out very efficient ($\mathcal{B}>1.9$) and very inefficient ($\mathcal{B}<0.2$) subhalo disruption at $95\%$ confidence (see \citealt{Nadler191203303} Sections 4.3 and 7.4).
    
    \item \emph{WDM half-mode mass} ($M_{\mathrm{hm}}$): The characteristic mass scale describing the suppression of the WDM SHMF as defined in Equations \ref{eq:wdm_shmf} and \ref{eq:mhm_mwdm}. Due to the population statistics of faint satellites corresponding to low-mass halos, MW satellite analyses have recently achieved upper limits on the half-mode mass that fall below the minimum halo mass associated with observed systems. In particular, for thermal relic WDM, \cite{Nadler200800022} infer $M_{\mathrm{hm}}<10^{7.5}\ M_{\mathrm{\odot}}$ ($m_{\mathrm{WDM}}>7.0\ \mathrm{keV}$) at $95\%$ confidence. \cite{Nadler200800022} scale this constraint by a factor of the maximum possible ratio of the MW halo mass relative to the average host halo mass of the two realistic MW zoom-in simulations used in the inference, which increases $M_{\mathrm{hm}}$ constraints by $\sim 25\%$ and yields a fiducial constraint of $M_{\mathrm{hm}}<10^{7.6}\ M_{\mathrm{\odot}}$ ($m_{\mathrm{WDM}}>6.5\ \mathrm{keV}$) at $95\%$ confidence. We discuss the role of the MW halo mass in detail in Section \ref{sec:systematics}.
    \end{enumerate}

\subsection{Constraints from Milky Way Satellite Observations}

\begin{figure}
\hspace{-15mm}
\includegraphics[scale=0.48,trim={-0.5cm 0 0 0}]{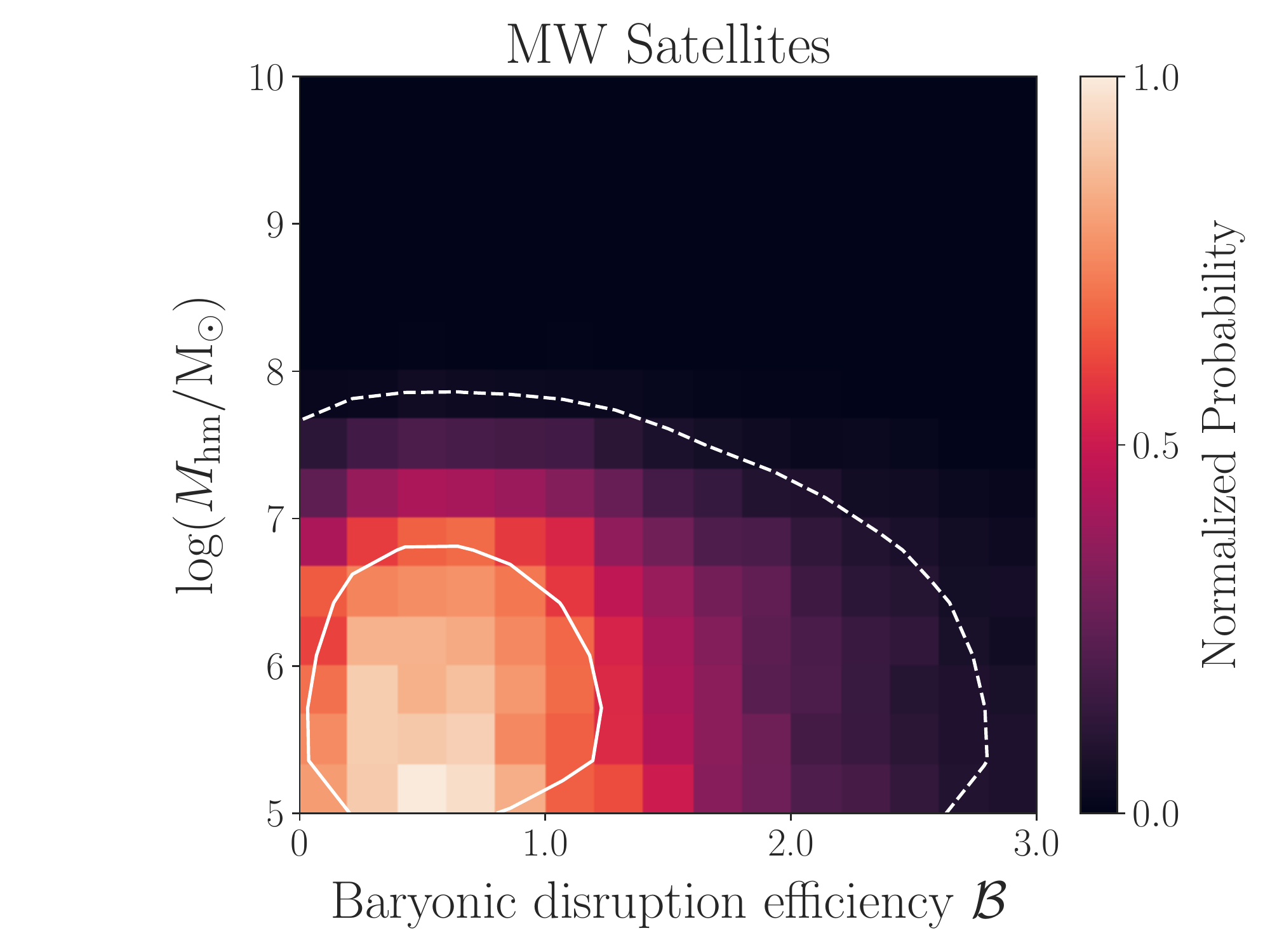}
\caption{Posterior distribution of WDM half-mode mass versus baryonic disruption efficiency from our MW satellite analysis. $\mathcal{B}=0$ corresponds to zero additional subhalo disruption relative to CDM, and larger values of $\mathcal{B}$ correspond to more efficient subhalo disruption due to baryons. The color map shows the probability density normalized to its maximum value in this parameter space. Solid (dashed) white lines indicate $1\sigma$ ($2\sigma$) contours for a two-dimensional Gaussian distribution.}
\label{fig:satellite_hist_raw}
\end{figure}

Here, we rerun the \cite{Nadler200800022} WDM MW satellite analysis using priors and a WDM SHMF parameterization chosen to match the \cite{Gilman190806983} lensing analysis, which allows us to self-consistently perform our multidimensional satellite--lensing probe combination. In particular, we rerun the MW satellite analysis adopting a uniform prior of $\mathcal{B}\sim \mathcal{U}(0,3)$, which ensures that we match the shape of the $\Sigma_{\mathrm{sub}}$ prior used in \cite{Gilman190806983} based on the linear relation between $\mathcal{B}$ and $\Sigma_{\mathrm{sub}}$ we derive in Section \ref{sec:b--sigma_sub}. The use of a uniform (rather than lognormal) prior on $\mathcal{B}$ weakens the upper and lower limits of the marginalized posterior from \cite{Nadler200800022} (i.e., $0.2<\mathcal{B}<1.9$) from $95\%$ to $68\%$ confidence constraints. In Appendix \ref{sec:priors}, we show that the choice of this prior does not significantly impact our joint WDM constraints.
    
We also use the WDM SHMF and the $\log M_{\mathrm{hm}}\sim \mathcal{U}(5,10)$ prior assumed in \cite{Gilman190806983}. The resulting marginalized posterior distribution yields $M_{\mathrm{hm}}<10^{7.4}\ M_{\mathrm{\odot}}$ ($m_{\mathrm{WDM}}>7.4\ \mathrm{keV}$) at $95\%$ confidence after MW host halo mass scaling, which is more constraining than the \cite{Nadler200800022} result despite using a slightly less suppressed WDM SHMF. This is caused by the change in the lower limit of our $\log M_{\mathrm{hm}}$ prior, which is two orders of magnitude lower than that adopted in \cite{Nadler200800022}. As described in Section \ref{sec:summarystats} the lower limit of the $M_{\mathrm{hm}}$ prior is arbitrary unless we assume that WDM physics manifests at a particular halo mass scale. Thus, we also quote likelihood ratios for both our independent and combined constraints. We find that $M_{\mathrm{hm}} = 10^{7.9}\ M_{\mathrm{\odot}}$ ($m_{\mathrm{WDM}} = 5.2\ \mathrm{keV}$) is disfavored relative to the peak of the marginalized posterior at $10^5\ M_{\mathrm{\odot}}$ with a 20:1 ratio, consistent with the \cite{Nadler200800022} result.
    
Figure \ref{fig:satellite_hist_raw} shows the posterior from our updated WDM fit to MW satellite data in the two-dimensional parameter space of $M_{\mathrm{hm}}$ versus $\mathcal{B}$, marginalized over seven other galaxy--halo connection parameters. In Figure \ref{fig:satellite_hist_raw} and subsequent plots, we do not scale parameters to account for MW host halo mass uncertainty unless explicitly noted. We reiterate that our MW satellite analysis only probes systems down to a peak halo mass threshold of $\sim 3\times 10^8\ M_{\mathrm{\odot}}$ at $95\%$ confidence and that $M_{\mathrm{hm}}$ constraints below this mass scale are driven by the population statistics of halos near the minimum observable halo mass. This is demonstrated in \citet[][see their Figure 1]{Nadler200800022}, which shows that the WDM model ruled out by MW satellites at $95\%$ confidence yields $\sim 25\%$ suppression in subhalo abundances relative to CDM at the minimum halo mass, which is about one order of magnitude larger than~$M_{\mathrm{hm}}$.

There is not a strong degeneracy between $\mathcal{B}$ and $M_{\mathrm{hm}}$ in Figure \ref{fig:satellite_hist_raw} because $\mathcal{B}$ models the disruptive effects of the MW disk, which suppresses the inner radial distribution of MW satellites in an approximately mass-independent fashion, while $M_{\mathrm{hm}}$ models the mass-dependent suppression of the projected SHMF caused by WDM free streaming. Figures \ref{fig:shmf} and \ref{fig:comparison_mvir} illustrate the effects of $\mathcal{B}$ on the projected SHMF and radial distribution of our MW-like simulations.

\section{Strong Lens Substructure Modeling}
\label{sec:lens}

Next, we describe the data and constraints from the \cite{Gilman190806983} quadruply lensed quasar flux ratio analysis our study is based on. Briefly, this analysis combines recent observations of the flux ratios and image positions from eight quadruply imaged quasars with a forward model for the dark matter substructure and line-of-sight halo populations to statistically infer the abundance and concentration of low-mass halos, which in turn constrains the WDM particle mass. Again, we refer the reader to specific sections of \cite{Gilman190806983} for modeling details throughout the following subsections.

\subsection{Strong Lensing Data}

\cite{Gilman190806983} analyze the narrow-line emission from six background quasars presented in \cite{Nierenberg190806344} and from two additional quasars presented in \cite{Nierenberg14021496,Nierenberg170105188}. These sources have a range of redshifts from $0.8\lesssim z_s \lesssim 3.7$, while the deflectors span redshifts of $0.2\lesssim z_d\lesssim 1$ and consist of massive elliptical galaxies. Priors on the masses of the deflector halos are estimated using the stellar mass--velocity dispersion relation derived for strong lens galaxies by \cite{Auger10072880}, and typically peak at $\sim 10^{13.3}\ M_{\mathrm{\odot}}$.
We note that \cite{Gilman190806983} excluded quads with main lensing galaxies that contain stellar disks from their analysis.

\subsection{Strong Lensing Model}

\begin{figure}
    \hspace{-3mm}
    \includegraphics[scale=0.42]{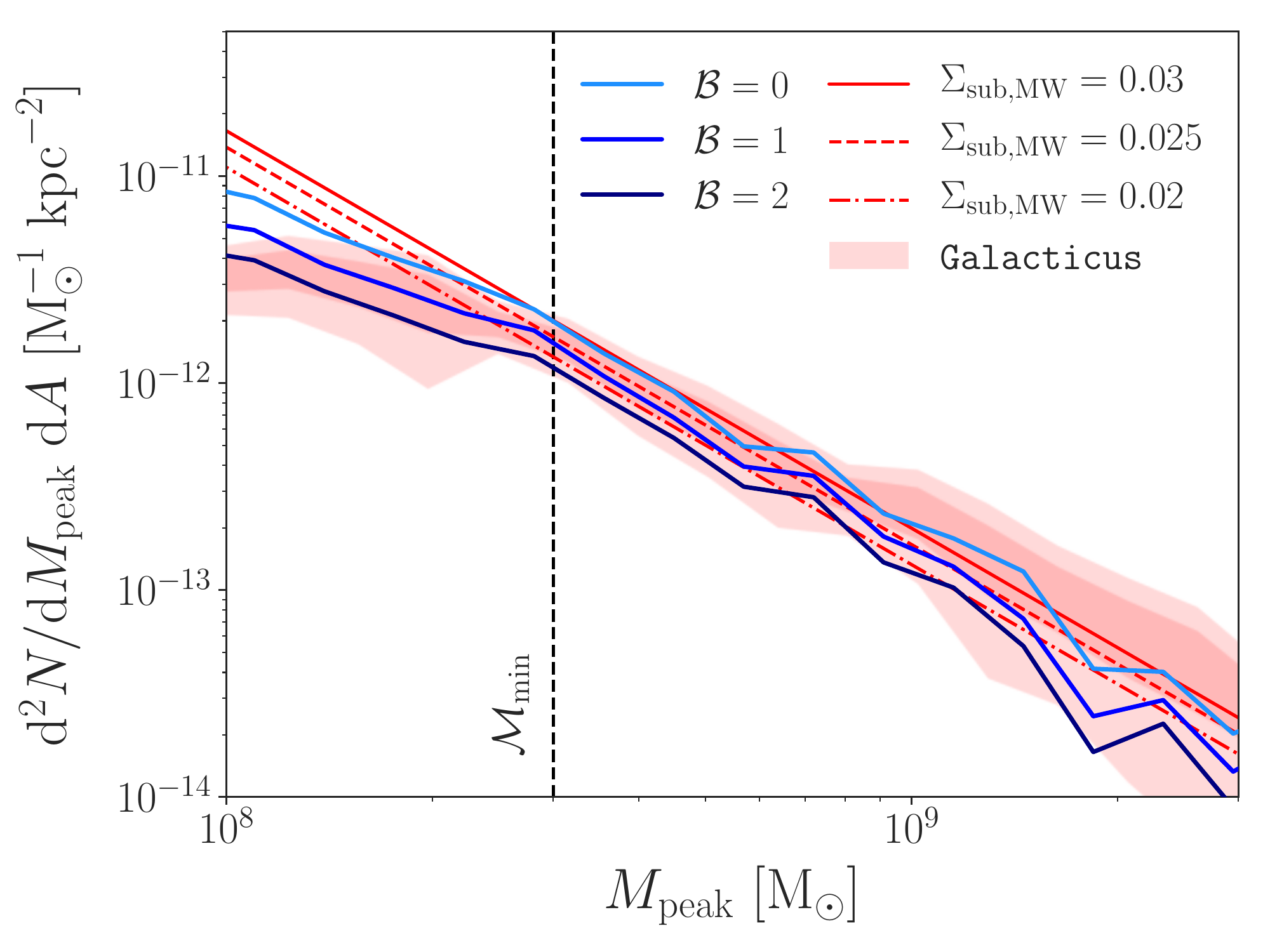}
    \caption{Projected SHMFs for MW-like host halos. Blue lines show the average results from the zoom-in simulations used in our MW satellite inference as a function of baryonic disruption efficiency $\mathcal{B}$ ($\mathcal{B}=0$ corresponds to CDM only and larger values of $\mathcal{B}$ correspond to more efficient subhalo disruption due to baryons). Red lines show our analytic SHMF (Equation \ref{eq:shmf_mw}) using the host halo mass and redshift scaling predicted by \texttt{Galacticus}, evaluated at the average halo mass of our MW-like simulations with a slope of $\alpha=-1.92$. $\Sigma_{\mathrm{sub,MW}}$ is chosen such that the SHMF amplitude matches our simulations at the subhalo mass corresponding to the faintest observed MW satellites, $\mathcal{M}_{\mathrm{min}}$ (dashed vertical line). Dark (light) red contours show $68\%$ ($95\%$) confidence intervals from \texttt{Galacticus} for host halos with characteristics matched to our MW-like simulations. We impose the resolution cuts described in Appendix \ref{sec:sim_details} on the simulation and \texttt{Galacticus} results.}
    \label{fig:shmf}
\end{figure}

The presence of small-scale structure in the lens and along the line of sight can perturb the magnified fluxes of unresolved quasar emission regions. The occurrence rate of the distribution of perturbed flux ratios between the multiple images is therefore sensitive to the underlying population of (potentially dark) subhalos within the host halo of the lens. Importantly, the strong lensing image position and flux ratio data described above are also sensitive to dark matter structure along the entire line of sight from the observer to the lensed quasars.

To perform the inference on the underlying subhalo and line-of-sight mass function population parameters, \cite{Gilman190806983} forward-model the quasar flux ratio with a large set of realizations of the small-scale lensing structure through a multiplane ray-tracing scheme, which accounts for the finite emitting source size and satisfies the astrometric constraints on the positions of the images. The likelihood for the individual lenses' population parameters was constructed with Approximate Bayesian Computation (ABC), and the joint posterior inference was performed by multiplying the individual likelihoods (see \citealt{Gilman190806983} Section 2).

Unlike in the case of subhalos, the line-of-sight dark matter structure is unaffected by tidal stripping and disruption. Thus, \cite{Gilman190806983} modeled the line-of-sight dark matter distribution using a Sheth--Tormen \citep{Sheth2001} mass function with the same WDM SHMF and concentration suppression factors described above for subhalos, along with a contribution from the two-halo term near the main deflector's host halo and an overall scaling factor that allows for uncertainty in the halo mass function amplitude of $20\%$ (see \citealt{Gilman190806983} Section 5.3).

\subsection{Constraints from Strong Lensing Observations}

The \cite{Gilman190806983} strong lensing analysis is consistent with CDM predictions for the slope and amplitude of the halo and subhalo mass functions. In particular, \cite{Gilman190806983} derive constraints on the SHMF slope that are consistent with $N$-body simulations and find that the line-of-sight contribution is consistent with Sheth--Tormen mass function predictions. Moreover, \cite{Gilman190902573} demonstrate that these data are compatible with standard predictions for the CDM mass--concentration relation while self-consistently modeling the effects of tidal stripping on subhalos.

\begin{figure*}[t]
    \hspace{-10mm}
    \includegraphics[scale=0.48]{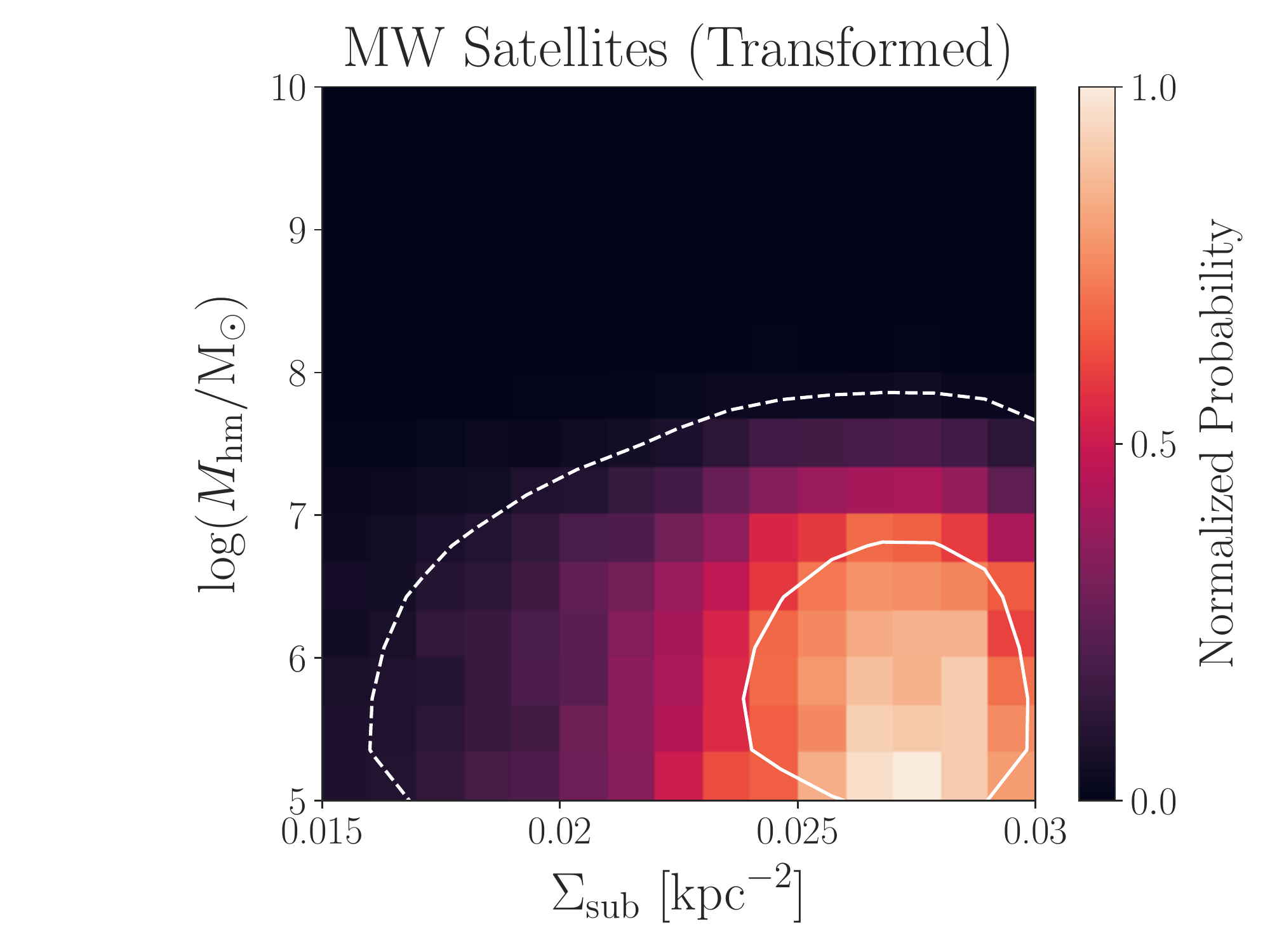}
    \includegraphics[scale=0.48]{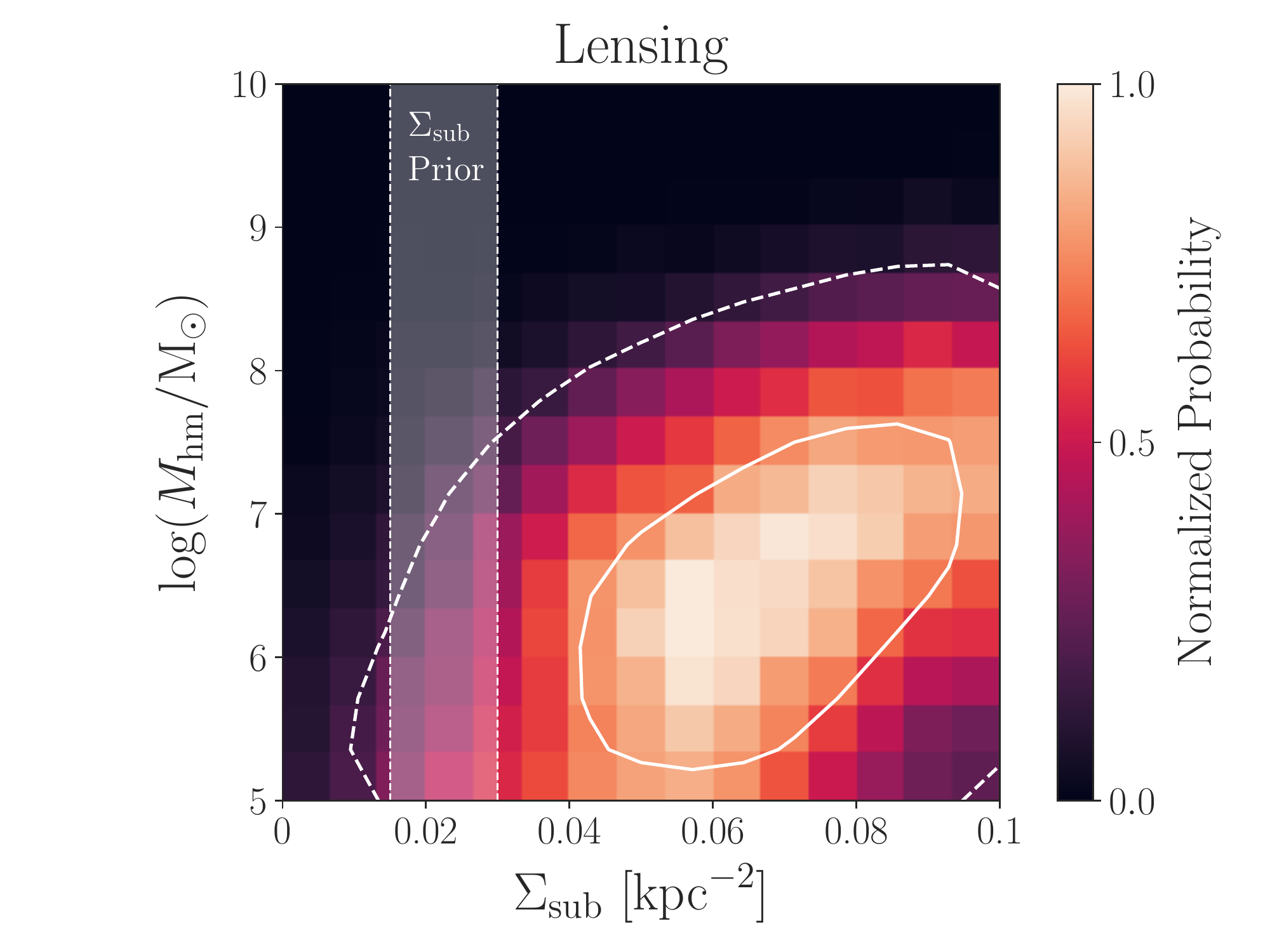}
    \caption{Left panel: posterior distribution of WDM half-mode mass versus projected subhalo number density at the strong lensing scale inferred from the MW satellite posterior, transformed according to the procedure in Section \ref{sec:b--sigma_sub}, with $q=1$ (i.e., for equally efficient subhalo disruption due to baryons at the MW and strong lensing host halo mass and redshift scales). Right panel: posterior distribution in the same parameter space from the \cite{Gilman190806983} strong lensing analysis. The vertical band labeled ``$\Sigma_{\mathrm{sub}}$ Prior'' shows the range of $\Sigma_{\mathrm{sub}}$ inferred from the MW satellite posterior in our fiducial joint analysis (i.e., $0.015\ \mathrm{kpc}^{-2}\leq \Sigma_{\mathrm{sub}} \leq 0.03\ \mathrm{kpc}^{-2}$). In both panels, color maps show the probability density normalized to its maximum value in each parameter space, and solid (dashed) white lines indicate $1\sigma$ ($2\sigma$) contours for a two-dimensional Gaussian distribution.}
    \label{fig:transformed_hist}
\end{figure*}

Here, we highlight the constraints from \cite{Gilman190806983} that enter our multidimensional MW satellite--lensing probe combination:
\begin{enumerate}[wide, label=({\emph{\roman*}}), itemsep=0pt]
\item \emph{Projected subhalo number density} ($\Sigma_{\mathrm{sub}}$): The amplitude of the projected SHMF defined in Equation \ref{eq:shmf_lensing}. \cite{Gilman190806983} place a lower limit of $\Sigma_{\mathrm{sub}}>8\times 10^{-3}\ \mathrm{kpc}^{-2}$ at $95\%$ confidence, which (given the lens sample) implies the presence of halos in the $10^6\ M_{\mathrm{\odot}}$--$10^9\ M_{\mathrm{\odot}}$ range. Lower values of $\Sigma_{\mathrm{sub}}$ do not yield sufficient perturbations to reproduce the observed flux ratios, and the \cite{Gilman190806983} analysis does not place an upper limit on $\Sigma_{\mathrm{sub}}$ within the prior range of $\Sigma_{\mathrm{sub}}\sim \mathcal{U}(0,0.1)\ \mathrm{kpc}^{-2}$ (see \citealt{Gilman190806983} Sections 3.2 and 6.2).
\item \emph{WDM half-mode mass} ($M_{\mathrm{hm}}$): The characteristic mass scale describing the suppression of the WDM SHMF defined in Equations \ref{eq:wdm_shmf} and \ref{eq:mhm_mwdm}. For thermal relic WDM, \cite{Gilman190806983} infer $M_{\mathrm{hm}}<10^{7.8}\ M_{\mathrm{\odot}}$ ($m_{\mathrm{WDM}}>5.6\ \mathrm{keV}$) at $95\%$ confidence. This constraint results from the fact that warmer models suppress the abundance and concentrations of low-mass halos that contribute to the lensing signal (see \citealt{Gilman190806983} Sections 3.4 and 6.2).

Here, we reanalyze the \cite{Gilman190806983} marginalized $M_{\mathrm{hm}}$ posterior using a slightly higher lower limit of the $\log M_{\mathrm{hm}}$ prior. We find $M_{\mathrm{hm}}<10^{8}\ M_{\mathrm{\odot}}$ ($m_{\mathrm{WDM}}>4.9\ \mathrm{keV}$) at $95\%$ confidence, which is slightly less constraining than the \cite{Gilman190806983} result. Again, we also calculate likelihood ratios due to the ambiguity of the $M_{\mathrm{hm}}$ prior and find that $M_{\mathrm{hm}}=10^{8.7}\ M_{\mathrm{\odot}}$ ($m_{\mathrm{WDM}}=3.0\ \mathrm{keV}$) is disfavored relative to the peak of the marginalized posterior at $10^{6.4}\ M_{\mathrm{\odot}}$ with a 20:1 ratio, consistent with \cite{Gilman190806983}.\footnote{We have resolved minor errors in the $M_{\mathrm{hm}}$--$m_{\mathrm{WDM}}$ conversion and likelihood ratios quoted in the original version of \cite{Gilman190806983}.}
\end{enumerate}

The right panel of Figure \ref{fig:transformed_hist} shows the posterior distribution from the fit to strong lensing data in \cite{Gilman190806983} in the two-dimensional parameter space of $M_{\mathrm{hm}}$ versus $\Sigma_{\mathrm{sub}}$, marginalized over the SHMF slope and line-of-sight mass function amplitude. There is a moderate degeneracy between $\Sigma_{\mathrm{sub}}$ and $M_{\mathrm{hm}}$, particularly at high values of $\Sigma_{\mathrm{sub}}$; in this regime, it is difficult to distinguish the coincident suppression of the WDM SHMF and mass--concentration relation relative to CDM from changes to the normalization of the CDM SHMF.


\section{Joint Analysis Methodology}
\label{sec:bridge}

Having described the data, models, and constraints that enter our joint analysis, we now describe our procedure for combining MW satellite and strong lensing constraints in a shared, multidimensional parameter space. In particular, we qualitatively outline our probe combination procedure (Section \ref{sec:outline}) and present our method for translating the subhalo disruption efficiency inferred from our MW satellite analysis to projected subhalo number density at the strong lensing scale (Section \ref{sec:b--sigma_sub}). We then describe the statistics of our probe combination (Section \ref{sec:probe_combination_stats}).

\subsection{Probe Combination Procedure}
\label{sec:outline}

\begin{figure}
    \hspace{-12mm}
    \includegraphics[scale=0.48]{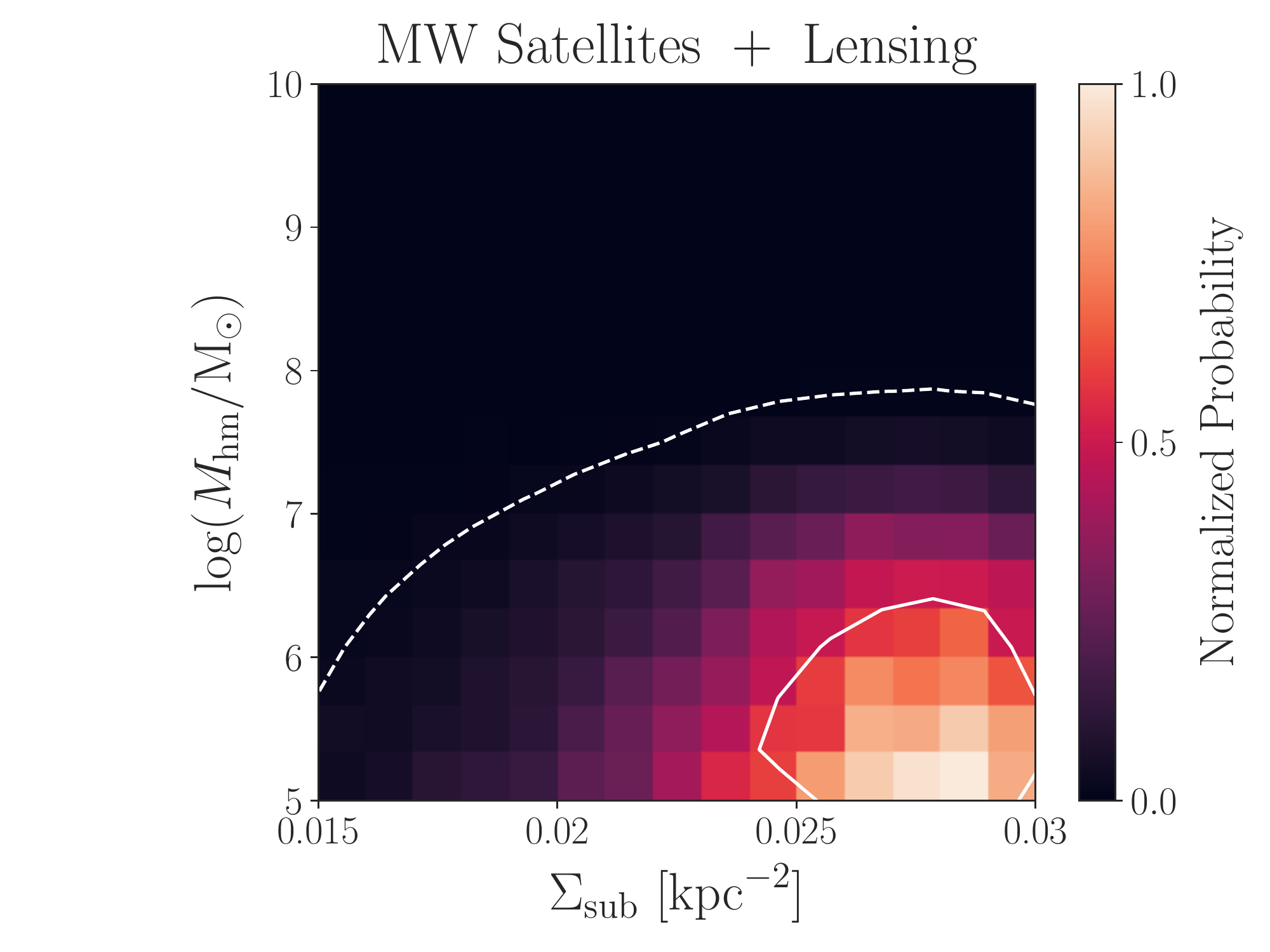}
    \caption{Joint marginal likelihood of WDM half-mode mass versus projected subhalo number density at the strong lensing scales from our combined MW satellite--strong lensing posterior, transformed according to the procedure in Section \ref{sec:b--sigma_sub}, with $q=1$. The color map shows the probability density normalized to its maximum value in this parameter space. Solid (dashed) white lines indicate $1\sigma$ ($2\sigma$) contours for a two-dimensional Gaussian distribution.}
    \label{fig:combined_hist}
\end{figure}

Our probe combination qualitatively proceeds as follows; these steps are described in detail in the following subsections:
\begin{enumerate}[wide, itemsep=0pt]
\item We compare \texttt{Galacticus} predictions for MW-mass halos to the projected SHMF inferred from the MW satellite population (Figure \ref{fig:shmf}) to construct a relation between the amplitude of the projected SHMF ($\Sigma_{\mathrm{sub,MW}}$) and the efficiency of subhalo disruption due to baryons in the MW ($\mathcal{B}$);
\item We use this relation to translate the $\mathcal{B}$--$M_{\mathrm{hm}}$ posterior from our MW satellite analysis (Figure \ref{fig:satellite_hist_raw}) into a $\Sigma_{\mathrm{sub,MW}}$--$M_{\mathrm{hm}}$ posterior distribution;
\item For a given value of the differential subhalo disruption efficiency $q$, we use Equation \ref{eq:q} to translate $\Sigma_{\mathrm{sub,MW}}$ to the strong lensing host halo mass and redshift regime, which yields a $\Sigma_{\mathrm{sub}}$--$M_{\mathrm{hm}}$ posterior from MW satellites that can be combined with the strong lensing posterior (Figure \ref{fig:transformed_hist});
\item We construct a joint $\Sigma_{\mathrm{sub}}$--$M_{\mathrm{hm}}$ likelihood by multiplying the MW satellite and strong lensing distributions according to the procedure in Section \ref{sec:probe_combination_stats} (Figure \ref{fig:combined_hist}).
\end{enumerate}

This method relies on several simplifying assumptions that could yield additional information if they are self-consistently addressed in a joint likelihood analysis using a model that simultaneously predicts the halo and subhalo distributions relevant for MW satellite and strong lensing analyses. We describe these areas for future work in Section \ref{sec:systematics}.

\subsection{Inferring $\Sigma_{\mathrm{sub}}$ from Milky Way Satellites}
\label{sec:b--sigma_sub}

To connect the host halo mass and redshift regimes probed by strong lensing and MW satellites, we first construct a relation between the subhalo disruption efficiency $\mathcal{B}$ inferred from our MW satellite analysis and the projected subhalo number density $\Sigma_{\mathrm{sub,MW}}$ predicted by evaluating \texttt{Galacticus} at the MW halo mass scale. We then translate $\Sigma_{\mathrm{sub,MW}}$ to $\Sigma_{\mathrm{sub}}$ at the strong lensing host halo mass and redshift scale using the dark matter substructure model described in Section \ref{sec:overview}.

Figure \ref{fig:shmf} shows the average $z=0$ projected SHMF from the two realistic zoom-in simulations used in our MW satellite analysis for our fiducial disruption model calibrated to hydrodynamic simulations (i.e., $\mathcal{B}=1$) and for bracketing values of the subhalo disruption efficiency (i.e., $\mathcal{B}=0$ and $\mathcal{B}=2$) that are ruled out at greater than $68\%$ confidence by MW satellite data as discussed in Section \ref{sec:mw}. To construct the projected SHMF predicted by our analytic substructure model, we use Equation \ref{eq:shmf_mw} with a slope of $\alpha=-1.92$ and with $\mathcal{F}_{\mathrm{CDM}}(M_{\mathrm{MW}},z_{\mathrm{MW}})$ evaluated at the mean virial mass of our simulated host halos, $M_{\mathrm{MW}}=1.4\times 10^{12}\ M_{\mathrm{\odot}}$, and $z_{\mathrm{MW}}=0$ to account for the CDM dependence on host halo mass and redshift. The zoom-in simulation predictions shown in these panels include both conservative resolution thresholds based on subhalos' peak and $z=0$ maximum circular velocity values as well as orphan subhalos (i.e., disrupted subhalos that are re-inserted into the simulation and analytically evolved until $z=0$) using the \cite{Nadler180905542} model; we provide additional details in Appendix \ref{sec:sim_details}. These choices allow for a more direct comparison to the semianalytically evolved subhalo populations predicted by \texttt{Galacticus}, which are less prone to artificial disruption \citep{VandenBosch171105276,VandenBosch180105427,Errani200107077}.

We construct a relation between $\Sigma_{\mathrm{sub,MW}}$ and $\mathcal{B}$ by matching our analytic prediction from Equation \ref{eq:shmf_mw} to the average projected SHMF predicted by our MW-like simulations as a function of $\mathcal{B}$, as illustrated in Figure \ref{fig:shmf}. In particular, we match the amplitudes of the projected SHMFs inferred from our MW satellite analysis and predicted by Equation \ref{eq:shmf_mw} at the minimum observable halo mass of $3.2\times 10^8\ M_{\mathrm{\odot}}$ and within a fixed radius of $300\ \mathrm{kpc}$ (roughly corresponding to the virial radius of the MW host halo), chosen to match the \cite{Nadler200800022} analysis. Our choice to match these SHMFs at the minimum halo mass is conservative because our MW satellite analysis is not sensitive to subhalos below this mass scale at $95\%$ confidence. We then translate $\Sigma_{\mathrm{sub,MW}}$ to the strong lensing regime using Equation~\ref{eq:q}, which yields
\begin{equation}
    \frac{\Sigma_{\mathrm{sub}}}{\mathrm{kpc}^{-2}} = \frac{0.03 - 0.005\mathcal{B}}{q}.\label{eq:b-relation}
\end{equation}
This relation allows us to transform the $\mathcal{B}$--$M_{\mathrm{hm}}$ MW satellite posterior from Figure \ref{fig:satellite_hist_raw} into the $\Sigma_{\mathrm{sub}}$--$M_{\mathrm{hm}}$ parameter space for a given value of $q$. Note that $\Sigma_{\mathrm{sub}}$ scales linearly with MW halo mass because it measures the projected SHMF amplitude within a fixed physical radius in our model.

The $\mathcal{B}$--$\Sigma_{\mathrm{sub,MW}}$ relation constructed above is only based on the amplitude of the projected SHMF from our two realistic MW-like simulations, measured at the minimum halo mass scale of $M_{\mathrm{peak}}=3\times 10^8\ M_{\mathrm{\odot}}$. In general, the procedure to infer $\Sigma_{\mathrm{sub,MW}}$ from an estimate of the MW SHMF should account for both host-to-host variation in the SHMF within the range of MW host halo properties allowed by observations and Poisson scatter in the SHMF given the range of subhalo masses probed by MW satellite measurements. Our SHMF matching procedure is intentionally simplistic because---as demonstrated in Appendix \ref{sec:scatter}---host-to-host and Poisson scatter in the projected SHMF near the minimum halo mass scale are both subleading uncertainties compared to the range of differential subhalo disruption efficiencies due to baryons that we explore. We therefore leave a statistically rigorous construction of the $\mathcal{B}$--$\Sigma_{\mathrm{sub}}$ relation to future work that propagates such uncertainties into the joint analysis at the likelihood level.

The result of the transformation in Equation \ref{eq:b-relation} is shown in the left panel of Figure \ref{fig:transformed_hist} for our fiducial model of $q=1$ (i.e., for equally  efficient subhalo disruption in the MW and strong lensing host halos). The typical $\Sigma_{\mathrm{sub}}$ values favored by the MW satellite posterior for this choice of $q$ are significantly smaller than the largest values allowed by the \cite{Gilman190806983} lensing analysis; we return to this point below.

The lack of degeneracy observed between $\mathcal{B}$ and $M_{\mathrm{hm}}$ in Figure \ref{fig:satellite_hist_raw}, which results from the joint constraining power of the MW satellite radial distribution and luminosity function for subhalos near the minimum observable halo mass, persists in the $\Sigma_{\mathrm{sub}}$--$M_{\mathrm{hm}}$ parameter space. On the other hand, strong lensing flux ratio statistics probe an integrated combination of subhalo masses and concentrations. The lensing analysis is currently less sensitive to subhalos in specific mass ranges than MW satellite population statistics and therefore exhibits a stronger $\Sigma_{\mathrm{sub}}$--$M_{\mathrm{hm}}$ degeneracy in Figure \ref{fig:transformed_hist}. However, because lensing measurements do not depend on the connection between subhalos and luminous matter, they can probe subhalos below the minimum observable halo mass.

In Figure \ref{fig:shmf} and Appendix \ref{sec:comparison}, we also compare our simulation results to the SHMF in terms of both peak and present-day subhalo mass and to the radial subhalo distribution predicted by \texttt{Galacticus} for 14 halos selected to match the characteristics of our MW-like simulations. In all cases, we apply the same cuts on peak and present-day subhalo maximum circular velocity when comparing \texttt{Galacticus} to our zoom-in simulations; the details of these resolution cuts and our \texttt{Galacticus} runs are described in Appendix \ref{sec:comparison}. Note that these \texttt{Galacticus} predictions should be compared to our $\mathcal{B}=0$ simulation results because the current \texttt{Galacticus} implementation does not model subhalo disruption due to central galaxies. The \texttt{Galacticus}-predicted SHMFs agree well with our simulations in terms of both peak and present-day subhalo mass, lending confidence to the choice of $\mathcal{F}_{\mathrm{CDM}}(M_{\mathrm{host}},z_{\mathrm{host}})$ that enters our calibration of the projected subhalo number density at the MW scale.

Because we construct a $\Sigma_{\mathrm{sub}}$--$\mathcal{B}$ relation based on our specific MW-like simulations, which have been shown to match the MW satellite population, we imposed several host halo selection criteria on the \texttt{Galacticus} runs used for the comparison above. These conditions include the existence of a realistic LMC analog system and a Gaia-Enceladus-like merger event. We emphasize that validating semianalytic halo and subhalo population predictions with self-consistent simulation suites of both MW-like and strong-lens-like halos is an important avenue for future work. As discussed in Section \ref{sec:systematics}, increasingly precise near-field observations and complementary data for strong lens systems will allow us to mitigate the impact of additional host halo properties including concentration (which is known to influence subhalo populations at fixed halo mass, e.g., \citealt{Zentner0411586,Zhu0601120,Ishiyama08120683,Mao150302637,Fielder180705180}) and the characteristics of the local dark matter environment in future joint analyses.

\subsection{Probe Combination Statistics}
\label{sec:probe_combination_stats}

Having placed the lensing and MW satellite posteriors on the same footing, we now proceed to combine them to construct a joint $\Sigma_{\mathrm{sub}}$--$M_{\mathrm{hm}}$ likelihood as follows. Formally, we write our joint MW satellite and strong lensing analysis as a combined Bayesian inference problem, 
\begin{flalign}
    P(\boldsymbol{\theta}\lvert \mathbf{D}) &\propto P(\mathbf{D}\lvert \boldsymbol{\theta})\times P(\boldsymbol{\theta})&\nonumber \\ &= P(\mathbf{D}_{\mathrm{MW}}\lvert \boldsymbol{\theta}_{\mathrm{MW}})\times P(\mathbf{D}_{\mathrm{lensing}}\lvert \boldsymbol{\theta}_{\mathrm{lensing}})\times P(\boldsymbol{\theta}),&
\end{flalign}
where $\boldsymbol{\theta}$ is the vector of parameters in both the lensing and satellite analyses including $q$ (where the shared parameters $\Sigma_{\mathrm{sub}}$ and $M_{\mathrm{hm}}$ only appear once), $\boldsymbol{\theta}_{\mathrm{MW}}$ ($\boldsymbol{\theta}_{\mathrm{lensing}}$) are the parameters entering the MW satellite (strong lensing) inference, $\mathbf{D}=[\mathbf{D}_{\mathrm{MW}},\mathbf{D}_{\mathrm{lensing}}]$ is the joint datavector, and $P(\boldsymbol{\theta})$ is the prior distribution over all model parameters.

Next, for a given value of $q\in [0.5,2]$, we marginalize over the independent parameters (i.e., the seven galaxy--halo connection parameters in the satellite analysis described in Section \ref{sec:mw} and the SHMF slope and line-of-sight contribution in the lensing analysis described in Section \ref{sec:lens}) to arrive at a combined $\Sigma_{\mathrm{sub}}$--$M_{\mathrm{hm}}$ posterior distribution,
\begin{flalign}
    P(\Sigma_{\mathrm{sub}},M_{\mathrm{hm}}\lvert \mathbf{D}) &= P(\Sigma_{\mathrm{sub}},M_{\mathrm{hm}}\lvert \mathbf{D}_{\mathrm{MW}}) \times P(\Sigma_{\mathrm{sub}},M_{\mathrm{hm}}\lvert \mathbf{D}_{\mathrm{lensing}})& \nonumber \\ &\propto P(\mathbf{D}\lvert \Sigma_{\mathrm{sub}},M_{\mathrm{hm}})\times P(\Sigma_{\mathrm{sub}},M_{\mathrm{hm}}).&\label{eq:bayes}
\end{flalign}
We assume independent priors for $\Sigma_{\mathrm{sub}}$ and $M_{\mathrm{hm}}$; in particular, use the prior distributions from \cite{Gilman190806983},
\begin{align}
    &P(\Sigma_{\mathrm{sub}}/\mathrm{kpc}^{-2}) = \mathcal{U}(0,0.1)& \\
    &P(\log (M_{\mathrm{hm}}/M_{\mathrm{\odot}})) = \mathcal{U}(5,10).&
\end{align}
As described in Sections \ref{sec:mw}--\ref{sec:lens}, we choose a lower limit of $\log (M_{\mathrm{hm}}/M_{\mathrm{\odot}})=5$ because models with even lower $M_{\mathrm{hm}}$ values are indistinguishable from CDM in both our MW satellite and strong lensing analyses. For simplicity, we repeat our analysis at several fixed values of $q$ rather than marginalizing over this parameter. We note that our WDM limits marginalized over $q$ are nearly identical to our fiducial ($q=1$) result in the absence of a well-motivated, nonuniform prior for $q$ based on hydrodynamic simulations (see Section \ref{sec:vary_q}).

\begin{figure*}
    \hspace{-2mm}
    \includegraphics[scale=0.435]{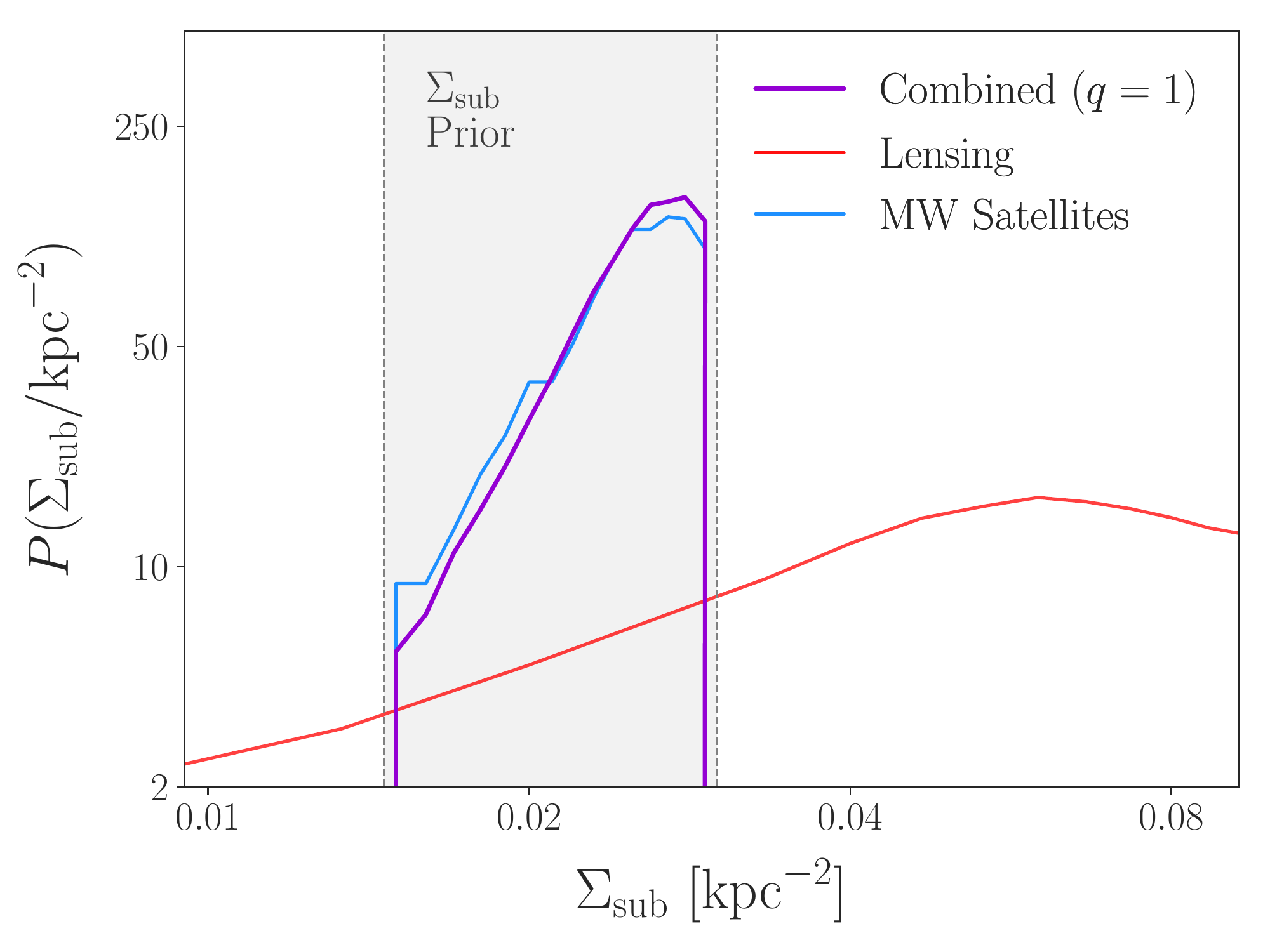}
    \hspace{2mm}
    \includegraphics[scale=0.435]{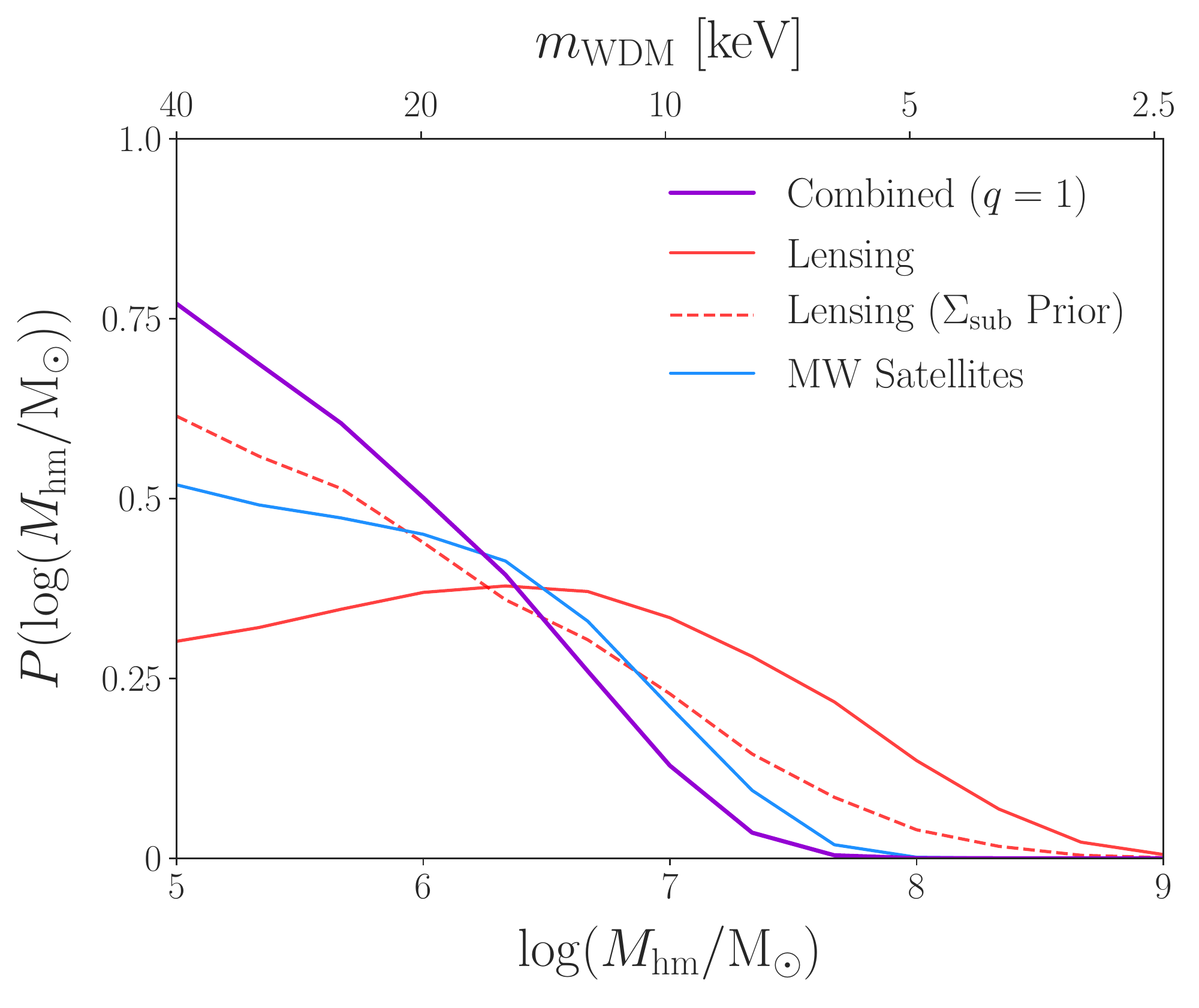}
    \caption{Marginal distributions from our joint MW satellite--strong lensing likelihood (Figure \ref{fig:combined_hist}) for projected subhalo number density at the strong lensing scale (left panel) and WDM half-mode mass (right panel), assuming equally efficient subhalo disruption due to baryons in the MW and strong lens systems ($q=1$). The marginalized MW satellite posterior is shown in blue, the marginalized strong lensing posterior is shown in red, and results obtained from our probe combination and marginalized over the remaining dimension are shown in purple. In the left panel, the vertical band labeled ``$\Sigma_{\mathrm{sub}}$ Prior'' shows the range of $\Sigma_{\mathrm{sub}}$ inferred from the MW satellite posterior in our fiducial joint analysis (i.e.,~$0.015\ \mathrm{kpc}^{-2}\leq \Sigma_{\mathrm{sub}} \leq 0.03\ \mathrm{kpc}^{-2}$, slightly offset from the posteriors for visual clarity), and the dashed red line on the right panel shows the lensing half-mode mass posterior restricted to this range of $\Sigma_{\mathrm{sub}}$ values.}
    \label{fig:1d_hists}
\end{figure*}

Based on Equation \ref{eq:b-relation}, our MW satellite inference only samples $\Sigma_{\mathrm{sub}}/\mathrm{kpc}^{-2}\in [0.015q^{-1},0.03q^{-1}] \in [0.0075,0.06]$ given our prior of $\mathcal{B}\sim \mathcal{U}(0,3)$ and our assumed range of $q\in [0.5,2]$ (note that $\mathcal{B}\geq 0$ by definition). Thus, our fiducial (i.e., $q=1$) analysis is restricted to the range $0.015\leq \Sigma_{\mathrm{sub}}/\mathrm{kpc}^{-2}\leq 0.03$, labeled ``$\Sigma_{\mathrm{sub}}$ Prior'' in Figures \ref{fig:transformed_hist} and \ref{fig:1d_hists}. This range is set by combining the MW satellite posterior and zoom-in simulations with our analytic SHMF prediction and is narrower than the $\Sigma_{\mathrm{sub}}$ range considered in \cite{Gilman190806983}, which did not enforce priors based on cosmological simulations. This difference limits the range of $\Sigma_{\mathrm{sub}}$ values from the lensing analysis relevant for our probe combination, but it does not formally affect our calculation of the marginal likelihood because the effective prior on $\Sigma_{\mathrm{sub}}$ from the MW satellite analysis is nevertheless uniform.

Thus, exploiting the fact that our priors are uniform in both $\Sigma_{\mathrm{sub}}$ and $\log M_{\mathrm{hm}}$, we arrive at a joint marginal likelihood for these quantities in terms of the marginalized MW satellite and strong lensing posteriors,
\begin{flalign}
    P(\mathbf{D}_{\mathrm{MW}},\mathbf{D}_{\mathrm{lensing}}\lvert \Sigma_{\mathrm{sub}},\log M_{\mathrm{hm}}) &\propto ~ P(\Sigma_{\mathrm{sub}},\log M_{\mathrm{hm}}\lvert \mathbf{D}_{\mathrm{MW}})&\nonumber \\ &\times P(\Sigma_{\mathrm{sub}},\log M_{\mathrm{hm}}\lvert \mathbf{D}_{\mathrm{lensing}}).&
\end{flalign}
This joint marginal likelihood is illustrated in Figure \ref{fig:combined_hist} and analyzed in Section \ref{sec:results}. Because the independently derived $M_{\mathrm{hm}}$ distributions are consistent (Section \ref{sec:q1}) and the $\Sigma_{\mathrm{sub}}$ distributions are only in mild tension (Section \ref{sec:SHMF_constraint}), we do not formally test for statistical consistency between the MW satellite and strong lensing analyses before constructing the joint likelihood.

Because we fix the slope of the projected SHMF in our analytic substructure model at $\alpha=-1.92$, our joint analysis is effectively performed at a thin slice in $\alpha$ of the \cite{Gilman190806983} posterior (we remind the reader that $\alpha$ is defined in a CDM context well above the cutoff scale). We do not expect this choice to significantly affect our results because $\alpha$ is not highly degenerate with the parameters of interest (i.e., $M_{\mathrm{hm}}$ and $\Sigma_{\mathrm{sub}}$) in the \cite{Gilman190806983} analysis. However, in Section \ref{sec:systematics} we emphasize the importance of jointly inferring the SHMF slope in future work, and we discuss the role of the remaining line-of-sight dimension of the \cite{Gilman190806983} posterior, modeled by the amplitude of the Sheth--Tormen halo mass function, that is marginalized over in our analysis.

\section{Results}
\label{sec:results}

We now present our probe combination results, which are summarized in Figures \ref{fig:combined_hist}--\ref{fig:1d_hists} and Table \ref{tab:q}. We first describe our conventions for calculating WDM constraints in Section~\ref{sec:summarystats}. We then describe our fiducial joint analysis results in Section~\ref{sec:q1}, and we explore the impact of varying the differential subhalo disruption efficiency due to baryons in Section~\ref{sec:vary_q} and our constraints on the projected SHMF amplitude in Section~\ref{sec:SHMF_constraint}.

\subsection{Conventions for WDM Constraints}
\label{sec:summarystats}

To quantify the WDM constraints corresponding to the joint likelihood derived in Section \ref{sec:probe_combination_stats}, we marginalize over the $\Sigma_{\mathrm{sub}}$ dimension and construct the following summary statistics from the marginal $M_{\mathrm{hm}}$ likelihood: 
\begin{enumerate}[wide, label=({\emph{\roman*}}), itemsep=0pt]
\item \emph{Confidence intervals}: defined as the range of parameter values enclosing a particular fraction of the integrated marginal likelihood. Following common practice in the WDM literature, we quote upper limits on $M_{\mathrm{hm}}$ and lower limits on $m_{\mathrm{WDM}}$ at $95\%$ confidence.
\item \emph{Marginal likelihood ratios}: defined as the parameter value at which the marginal likelihood probability density falls to a particular fraction of its peak value. Following \cite{Gilman190806983}, we quote the $M_{\mathrm{hm}}$ and $m_{\mathrm{WDM}}$ values disfavored with 20:1 marginal likelihood ratios.
\end{enumerate}

\begin{deluxetable*}{l c c c c}[t!]
\centering
\tabletypesize{\footnotesize}
\tablewidth{\columnwidth}
\tablecaption{\label{tab:q} $95\%$ Confidence and 20:1 Likelihood Ratio Upper Limits on $M_{\mathrm{hm}}$ and Corresponding Lower Limits on $m_{\mathrm{WDM}}$ for Our Multidimensional Probe Combination for Various Differential Subhalo Disruption Efficiency Values $q$, and for an Analysis That Combines the Fully Marginalized One-dimensional $M_{\mathrm{hm}}$ Distributions.}
\tablehead{\colhead{} & \colhead{One-dimensional $M_{\mathrm{hm}}$ Distributions} & \colhead{$q=0.5$} & \colhead{$q=1$} & \colhead{$q=2$}}
\startdata
$95\%$ confidence level $M_{\mathrm{hm}}\ [M_{\mathrm{\odot}}]$ & $10^{7.2}$ & $10^{7.1}$ & $10^{7.0}$ & $10^{6.9}$ \\
$95\%$ confidence level $m_{\mathrm{WDM}}\ [\mathrm{keV}]$ &  $8.4$ & $9.1$ & $9.7$ & $10.4$ \\
\hline
20:1 likelihood ratio $M_{\mathrm{hm}}\ [M_{\mathrm{\odot}}]$ & $10^{7.7}$&  $10^{7.6}$ & $10^{7.4}$ & $10^{7.3}$  \\
20:1 likelihood ratio $m_{\mathrm{WDM}}\ [\mathrm{keV}]$ & $6.0$ & $6.4$ & $7.4$ & $7.9$  \\
\enddata
{\footnotesize \tablecomments{$q=0.5$ corresponds to twice as efficient subhalo disruption due to baryons in the MW relative to strong lenses, $q=1$ (our fiducial model) corresponds to equally efficient subhalo disruption due to baryons, and $q=2$ corresponds to twice as efficient subhalo disruption due to baryons in strong lenses.}\vspace{-5mm}}
\end{deluxetable*}

Although confidence intervals capture more information about the shape of the probability density and are commonly quoted in the WDM literature (e.g., \citealt{Viel1308804,Irsic179602}), they depend on the arbitrary choice of a lower limit on the $M_{\mathrm{hm}}$ prior (or equivalently, an upper limit on the $m_{\mathrm{WDM}}$ prior) as noted above. In particular, small-scale structure data are currently consistent with CDM and therefore yield one-sided limits on $M_{\mathrm{hm}}$ or $m_{\mathrm{WDM}}$; without assuming a preferred scale for a small-scale structure cutoff due to WDM (or other non-CDM) physics, this makes the lower limit of the $M_{\mathrm{hm}}$ prior arbitrary. This situation motivated several authors (e.g., \citealt{Enzi201013802,Gilman190806983}) to quote alternative summary statistics including marginal likelihood ratios that are less dependent on the choice of $M_{\mathrm{hm}}$ prior, and we follow this practice here. Similarly, we follow both \cite{Gilman190806983} and \citealt{Nadler200800022} by adopting a logarithmic prior on $M_{\mathrm{hm}}$ because any other choice would not be invariant to rescaling $m_{\mathrm{WDM}}$ (e.g., see the discussion in \citealt{Jethwa161207834}).

\subsection{Fiducial WDM Constraints}
\label{sec:q1}

We now present the results of our joint analysis for our fiducial subhalo disruption efficiency model of $q=1$, which assumes equally efficient subhalo disruption due to baryons in the MW and in strong lens host halos, which is broadly compatible with the results of hydrodynamic simulations (see Section \ref{sec:q_description}). The combined $\Sigma_{\mathrm{sub}}$--$M_{\mathrm{hm}}$ marginal likelihood is shown in Figure \ref{fig:combined_hist} and the corresponding one-dimensional marginalized likelihoods for $\Sigma_{\mathrm{sub}}$ and $M_{\mathrm{hm}}$ are shown in Figure \ref{fig:1d_hists}. The joint marginal likelihood retains the shape of the $\Sigma_{\mathrm{sub}}$--$M_{\mathrm{hm}}$ distribution from the transformed MW satellite posterior and from the lensing analysis limited to the range of $\Sigma_{\mathrm{sub}}$ inferred from our MW satellite analysis according to the procedure in Section \ref{sec:b--sigma_sub}. Moreover, the joint marginal likelihood visibly prefers lower values of $M_{\mathrm{hm}}$ than either posterior alone, demonstrating the unique constraining power accessible when combining independent small-scale structure probes in a multidimensional parameter space.

Consistent with these qualitative aspects of the joint $\Sigma_{\mathrm{sub}}$--$M_{\mathrm{hm}}$ likelihood, the upper limit of the marginal $M_{\mathrm{hm}}$ likelihood shown in the right panel of Figure \ref{fig:1d_hists} is noticeably lower than either of the individual constraints from MW satellites or strong lensing. Quantitatively, the upper limit on $M_{\mathrm{hm}}$ from our joint analysis improves upon those set by the MW satellite and strong lensing analyses individually by \textcolor{black}{$\sim 60\%$}, leading to a \textcolor{black}{$\sim 30\%$} increase in the strength of the lower limit on $m_{\mathrm{WDM}}$. Specifically, the $95\%$ confidence limit of \textcolor{black}{$M_{\mathrm{hm}}<10^{7.4}\ M_{\mathrm{\odot}}$} (\textcolor{black}{$m_{\mathrm{WDM}}> 7.4\ \mathrm{keV}$}) from our MW satellite analysis improves to \textcolor{black}{$M_{\mathrm{hm}}<10^{7.0}\ M_{\mathrm{\odot}}$} (\textcolor{black}{$m_{\mathrm{WDM}}> 9.7\ \mathrm{keV}$}). We find a similar level improvement in terms of likelihood ratios, with \textcolor{black}{$M_{\mathrm{hm}}=10^{7.4}\ M_{\mathrm{\odot}}$} (\textcolor{black}{$m_{\mathrm{WDM}}=7.4\ \mathrm{keV}$}) ruled out at 20:1 relative to the peak of the marginal likelihood at the lower limit of the prior at $10^{5}\ M_{\mathrm{\odot}}$.

To derive these limits, we conservatively increased the $M_{\mathrm{hm}}$ values returned by our joint analysis by a factor of $\sim 25\%$ to account for the maximum mass of the MW halo relative to the average host halo masses of our zoom-in simulations, following \cite{Nadler200800022}. As demonstrated in the following subsection, propagating the MW halo mass uncertainty into the $\Sigma_{\mathrm{sub}}$ dimension would have a negligible impact on the results compared to uncertainties in the efficiency of subhalo disruption due to baryons, so we do not perform this scaling for simplicity. 

Our fiducial constraint of $m_{\mathrm{WDM}}> 9.7\ \mathrm{keV}$ at $95\%$ confidence is one of the most stringent limits on the WDM particle mass set by small-scale structure observations to date. Moreover, it is set using only existing strong lensing and MW satellite measurements, underscoring the importance of unified, multidimensional small-scale structure analyses as the corresponding measurements continue to improve. Joint model-building efforts that further incorporate Ly$\alpha$ forest \citep{Viel1308804,Irsic179602} and stellar stream \citep{Banik191102662} constraints while retaining the unique information provided by each probe will therefore be particularly fruitful.

\subsection{Impact of the Differential Subhalo Disruption Efficiency Due to Baryons}
\label{sec:vary_q}

\begin{figure*}[t!]
    \hspace{-2mm}
    \includegraphics[scale=0.435]{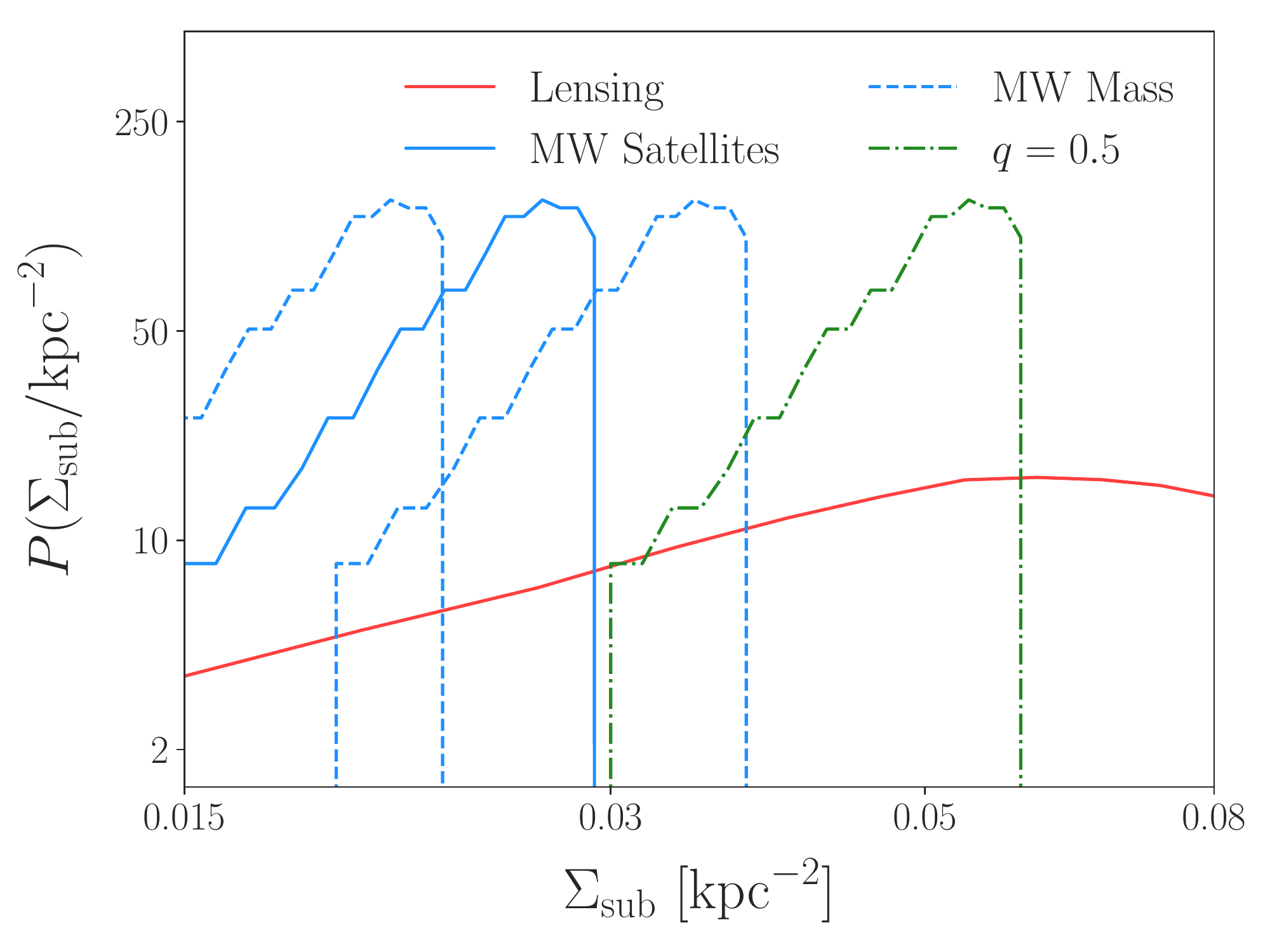}
    \hspace{2mm}
    \includegraphics[scale=0.435]{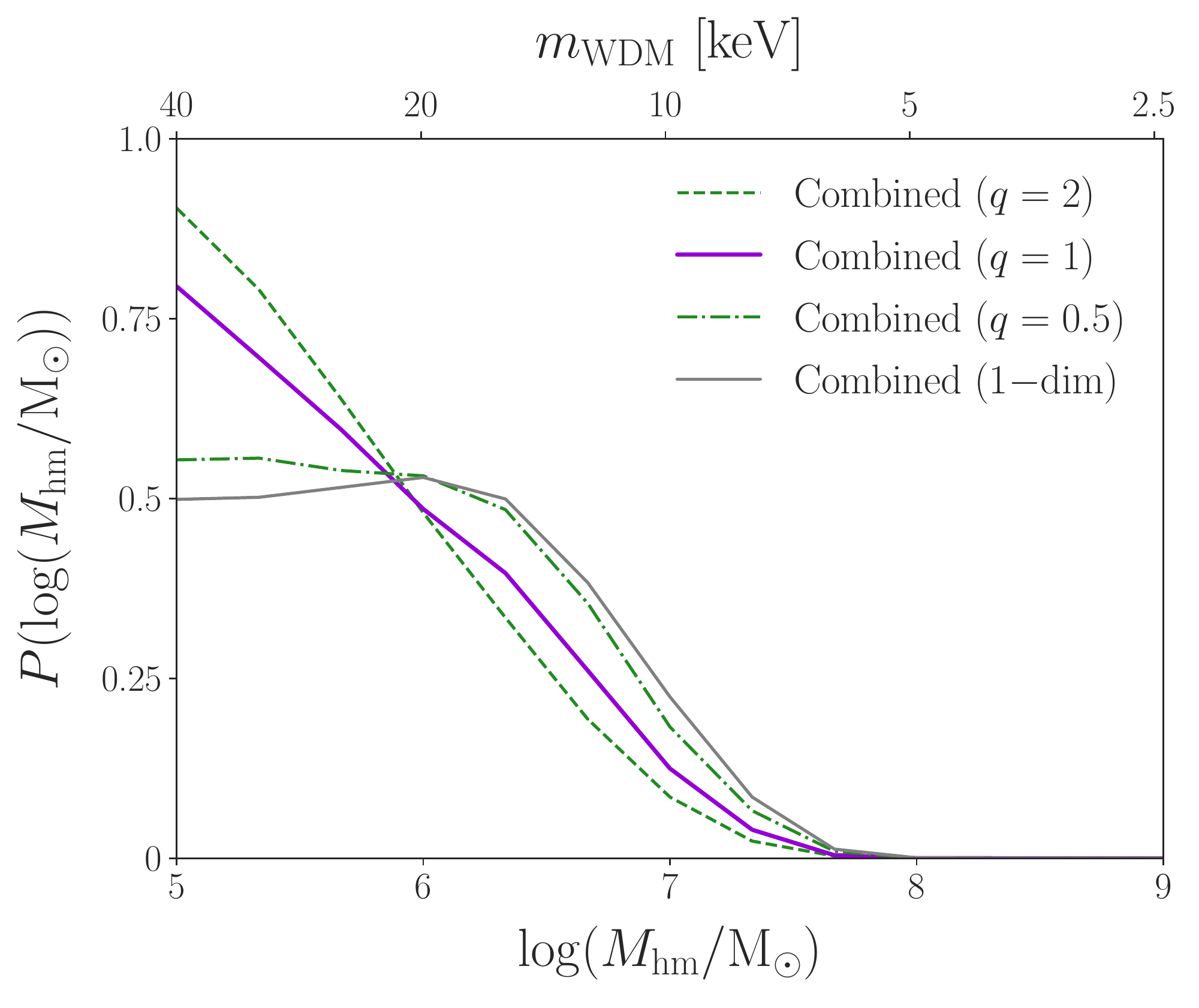}
    \caption{Left panel: the impact of systematics on the marginalized one-dimensional posterior distributions of projected subhalo number density at the strong lensing scale. The marginalized posterior distribution from our MW satellite analysis is shown in blue, the marginalized strong lensing posterior is shown in red, the dashed blue distributions indicate additional uncertainty in our MW satellite inference due to the mass of the MW halo, and the dotted--dashed green distribution illustrates the effects of systematic uncertainty in the differential efficiency of subhalo disruption due to baryons at the MW and strong lensing host halo scales. Right panel: joint marginal likelihood of WDM half-mode mass for our MW satellites plus strong lensing probe combination. Joint likelihoods are shown for equally efficient subhalo disruption in the MW and strong lens host halo mass and redshift regimes ($q=1$, purple), twice as efficient disruption due to baryons in the MW relative to strong lens halos ($q=0.5$, dotted--dashed green), and twice as efficient disruption in strong lens halos relative to the MW ($q=2$, dashed green). The gray distribution shows the result of combining the fully marginalized one-dimensional $M_{\mathrm{hm}}$ posteriors derived from strong lensing and MW satellites.}
    \label{fig:1d_hists_systematics}
\end{figure*}

We now explore the impact of the differential efficiency of subhalo disruption due to baryons on our WDM constraints. Table \ref{tab:q} lists the $M_{\mathrm{hm}}$ and $m_{\mathrm{WDM}}$ $95\%$ confidence level and 20:1 likelihood ratio limits for $q=0.5$, $1$, and $2$, and the right panel of Figure \ref{fig:1d_hists_systematics} shows the corresponding joint marginal likelihoods. In Table \ref{tab:q} and Figure \ref{fig:1d_hists_systematics}, we also show the result of combining the fully marginalized one-dimensional $M_{\mathrm{hm}}$ posteriors from our MW satellite and strong lensing analyses.

As demonstrated in the right panel of Figure \ref{fig:1d_hists_systematics}, the joint marginal likelihoods for $M_{\mathrm{hm}}$ become increasingly constraining as $q$ increases. This is due to the fact that the transformed $\Sigma_{\mathrm{sub}}$--$M_{\mathrm{hm}}$ posterior distribution from MW satellites (Figure \ref{fig:transformed_hist} left panel) breaks the degeneracy between these parameters present in the strong lensing posterior. In particular, larger values of $q$ correspond to more efficient subhalo disruption in strong lens host halos relative to the MW and yield lower inferred values of $\Sigma_{\mathrm{sub}}$ at the strong lensing scale according to Equation \ref{eq:b-relation}. This shifts the region of two-dimensional parameter space in which we multiply the MW satellite and strong lensing posteriors toward lower values of $\Sigma_{\mathrm{sub}}$. Thus, because the low-$\Sigma_{\mathrm{sub}}$ region of the lensing posterior does not allow for large values of $M_{\mathrm{hm}}$, larger values of $q$ yield more stringent joint $M_{\mathrm{hm}}$ constraints (and vice versa for smaller values of $q$). Indeed, as shown in the right panel of Figure \ref{fig:1d_hists}, restricting the strong lensing posterior to the range of $\Sigma_{\mathrm{sub}}$ inferred from our MW satellite analysis for $q=1$ significantly strengthens the $M_{\mathrm{hm}}$ constraint set by lensing alone.

Despite the qualitative effects of varying the differential subhalo disruption efficiency described above, varying $q$ within a reasonably broad range only impacts the results of our probe combination at the $\sim 10\%$ level in terms of $m_{\mathrm{WDM}}$. As discussed in Section \ref{sec:systematics}, the differential efficiency of subhalo disruption due to baryons is one of several systematics that impact our probe combination at this level, all of which must be controlled in a joint modeling framework to claim a detection of non-CDM physics at the corresponding level of precision. Figure \ref{fig:1d_hists_systematics} and Table~\ref{tab:q} demonstrate that combining the MW satellite and strong lensing posteriors with any value of $q$---i.e., performing the combination in multiple dimensions---is more constraining than combining the fully marginalized $M_{\mathrm{hm}}$ posteriors, as expected.

\subsection{Projected Subhalo Number Density Constraints}
\label{sec:SHMF_constraint}

As demonstrated in the left panel of Figure \ref{fig:1d_hists}, the marginalized posterior for $\Sigma_{\mathrm{sub}}$ from strong lensing accommodates significantly larger values than we infer from our MW satellite analysis. In particular, our fiducial joint analysis yields a marginalized posterior distribution from MW satellites that peaks at $\Sigma_{\mathrm{sub}}\approx 0.025\ \mathrm{kpc}^{-2}$; moreover, $\Sigma_{\mathrm{sub}}> 0.03\ \mathrm{kpc}^{-2}$ is not sampled because these projected SHMF amplitudes are larger than the average of our MW-like simulations.\footnote{The upper limit of this prior increases to $\Sigma_{\mathrm{sub}} \sim 0.04\ \mathrm{kpc}^{-2}$ when accounting for uncertainties in the mass of the MW host halo, which is still much lower than the largest $\Sigma_{\mathrm{sub}}$ inferred in \cite{Gilman190806983}.} Meanwhile, $\Sigma_{\mathrm{sub}}$ values in this range are disfavored in the lensing posterior relative to its mild peak at $\Sigma_{\mathrm{sub}}\approx 0.067\ \mathrm{kpc}^{-2}$ by a ratio of $\sim$ 2:1. Although this is not a significant tension, it is worth exploring in future work that places $\Sigma_{\mathrm{sub}}$ constraints at various host mass and redshift scales in the context of expectations from cosmological simulations. For example, \cite{Lazar201203958} identify potentially significant contributions from backsplash halo populations near strong lenses beyond those captured by the two-halo term used in \cite{Gilman190806983}, which (if modeled) may lower the inferred range of $\Sigma_{\mathrm{sub}}$ and strengthen the corresponding WDM constraints. Furthermore, there are potential differences between the surviving subhalo populations inferred from our MW satellite and strong lensing analyses caused by tidal stripping, although heavily stripped halos do not dominate the signal in either case. Thus, although it is unlikely because the subhalos that contribute to strong lensing flux ratio statistics are usually tidally truncated well outside of their NFW scale radius \citep{Gilman190806983,Minor201110629}, a careful analysis of whether these systems can be stripped severely enough such that their luminous content is affected warrants detailed investigation in future work.

In the left panel of Figure \ref{fig:1d_hists_systematics}, we show how the $\Sigma_{\mathrm{sub}}$ posterior from our MW satellite analysis shifts as a function of both the differential subhalo disruption efficiency due to baryons, $q$, and the MW halo mass, where we use the MW host halo mass uncertainties discussed in Section \ref{sec:mw} and assume that $\Sigma_{\mathrm{sub}}\propto M_{\mathrm{MW}}$. The $\Sigma_{\mathrm{sub}}$ distribution from MW satellites is clearly sensitive to both of these systematic uncertainties, which we discuss further in Section \ref{sec:discussion}. Because varying $q$ changes the inferred $\Sigma_{\mathrm{sub}}$ distribution in the strong lens halo mass and redshift regime, this quantity can potentially be constrained as the precision of $\Sigma_{\mathrm{sub}}$ constraints from strong lensing increases. Although we do not attempt to constrain $q$ here, we note that our results disfavor simultaneously high MW halo mass and low subhalo disruption efficiency due to baryons in the MW relative to strong lens host halos, which is physically reasonable.

\section{Discussion}
\label{sec:discussion}

We now place the WDM and SHMF constraints from our MW satellite--strong lensing probe combination in context by discussing key systematics (Section \ref{sec:systematics}) and comparing our study to other recent analyses (Section \ref{sec:study}).

\subsection{Systematics}
\label{sec:systematics}

The analysis presented above casts MW satellite and strong lensing constraints in a shared, multidimensional parameter space for the first time. We emphasize that our WDM constraints (Section \ref{sec:q1}) are conservative due to our broad priors on key systematics and are robust to the modeling uncertainties directly addressed in the joint analysis at the $\sim 10\%$ level (Section \ref{sec:vary_q}). Our work therefore provides important foundations for more detailed modeling frameworks that simultaneously constrain MW satellite and lensing observables at the likelihood level.

Nevertheless, our analysis makes several simplifying assumptions that circumvent a joint likelihood analysis. We regard these as crucial areas for future model-building work in preparation for next-generation facilities and surveys, both for the MW satellite--strong lensing probe combination presented here and to further combine these probes with analyses of stellar stream perturbations, the Ly$\alpha$ forest, and any other novel probes of small-scale structure. In general, joint small-scale structure constraints may be sensitive to additional ``nuisance parameters'' distinct from those governing non-CDM physics, which must be simultaneously measured to robustly claim evidence for a deviation from CDM. This underscores the importance of our multidimensional approach and of the following systematics, which we plan to build a joint model to simultaneously infer in future work.

\emph{SHMF slope}. We assume a particular value of the SHMF slope $\alpha$ when constructing the $\mathcal{B}$--$\Sigma_{\mathrm{sub}}$ relation in Section \ref{sec:bridge}, thereby taking a thin slice through this dimension of the posterior from \cite{Gilman190806983}. Although current MW satellite analyses do not strongly constrain the SHMF slope, future constraints from the MW satellite population probed by LSST may be sensitive to this quantity due to excellent observational sensitivity at the faint end of the satellite luminosity function throughout the MW virial radius (e.g., \citealt{Ivezic08052366,Hargis14074470,Drlica-Wagner190201055}). Meanwhile, the \cite{Gilman190806983} strong lensing analysis already mildly constrains the SHMF slope, and this sensitivity will drastically increase with larger lens samples. Exploiting all of these data will require self-consistent suites of high-resolution simulations of both MW-like systems (including realistic LMC analogs) and strong-lens-like systems, which we are currently developing. Few such high-resolution zoom-in simulations at the group-mass scale have been performed, and these are particularly valuable to validate the predictions of semianalytic models like \texttt{Galacticus} used to inform strong lens substructure models. These studies must be coupled with detailed models for the impact of baryonic physics on small-scale dark matter structure because it is expected to significantly affect both the amplitude and slope of SHMF at low halo masses \citep{Benson191104579}.

\emph{Line-of-sight halo mass function}. We marginalized over the amplitude of the line-of-sight halo mass function in our probe combination, noting that the \cite{Gilman190806983} lensing analysis our work is based on does not constrain this quantity within a broad prior range of $\pm 20\%$ relative to the mean Sheth--Tormen prediction. However, detailed zoom-in simulations of strong lens analogs coupled with realizations from cosmological simulations of the line-of-sight halo populations may provide more informative theoretical priors that---combined with upcoming strong lens discoveries and follow-up imaging and spectroscopy---will yield more decisive differential measurements of the line-of-sight and substructure contributions to the lensing signal (see \citealt{Lazar201203958} for a recent discussion). This will ultimately allow $\Sigma_{\mathrm{sub}}$ to be measured more precisely, breaking degeneracies with WDM physics and facilitating a more direct combination with MW satellite data.

\emph{Subhalo disruption efficiency due to baryons}. We combined MW satellite and strong lensing constraints at fixed values of the differential subhalo disruption efficiency due to baryons, $q$. Although $q$ does not significantly affect the joint WDM limits presented here (Section \ref{sec:results}), this quantity represents a key systematic that must be addressed in dedicated modeling work. In particular, it will be fruitful to analyze samples of hydrodynamic simulations at the MW and group-mass scales to refine subhalo disruption models that can be applied to larger simulation suites efficiently (e.g., \citealt{Nadler171204467}). Constructing a physically motivated model for the differential efficiency of subhalo disruption due to baryons in strong lens systems and the MW will again allow for more informative theoretical priors in joint analyses, enabling robust constraints on deviations from CDM predictions.

\emph{Milky Way and strong lens host halo properties}. The mass of the MW halo remains a key systematic for the interpretation of MW satellite measurements in terms of the underlying SHMF which then propagates into joint small-scale structure constraints. The MW halo mass is a particularly important nuisance parameter for setting non-CDM constraints because the (lack of a) turnover in low-mass subhalo abundances is inferred from the SHMF corresponding to MW satellite observations, while the SHMF amplitude scales linearly with host halo mass. In our analysis, uncertainty in the MW halo mass significantly affects our $M_{\mathrm{hm}}$ and $m_{\mathrm{WDM}}$ constraints, and we currently take a conservative approach to marginalize over this dependence. Forthcoming Gaia data releases will increase the precision of MW halo mass measurements, and combining detailed simulation suites of MW-like halos spanning the inferred mass range with next-generation observations of the MW satellite population will allow us to derive joint constraints on the MW halo mass and SHMF (e.g., see \citealt{Newton201108865}).

Meanwhile, strong lensing measurements are less sensitive to host halo mass uncertainty because they probe both the SHMF, small-scale structure along the line of sight, and the concentrations of low-mass halos and subhalos. Nonetheless, the details of strong lens host halo selection functions are relatively unexplored (e.g., see \citealt{Sonnenfeld14101881}), and will be better quantified using a variety of data including weak lensing and satellite velocity dispersion measurements. These efforts will lead to more precise constraints on the masses, secondary properties, and environments of strong lens host halos, further mitigating key theoretical uncertainties in forward models of strong lensing data.

\subsection{Comparison to Recent Studies}
\label{sec:study}

\cite{Enzi201013802} recently presented a joint analysis of small-scale structure probes including MW satellite galaxies and gravitational imaging, with several distinct assumptions underlying the individual and joint modeling of these probes relative to our work. Here, we discuss the most important aspects of our individual models for MW satellites and strong lensing flux ratio statistics as well as our probe combination procedure relative to the \cite{Enzi201013802} study.

For MW satellites, the \cite{Nadler200800022} study upon which we base our analysis explicitly includes realistic LMC analog systems in the simulations used to perform the inference. This allows \cite{Nadler200800022} to use the entire population of observed MW satellite galaxies---and particularly those within and near the DES footprint---without down-weighting systems based on the probability they are associated with the LMC, strengthening our dark matter constraints relative to the \cite{NewtonMNRAS,Newton201108865} MW satellite analyses that the \cite{Enzi201013802} joint constraints are based on. In addition, unlike \cite{NewtonMNRAS,Newton201108865}, we follow \cite{Nadler200800022} by using the newest and most precise versions of DES and PS1 observational selection functions from \cite{Drlica-Wagner191203302}. Importantly, these selection functions depend on satellite galaxy size, which is a crucial driver of satellite detectability that directly informs the translation from MW satellite observations to the underlying SHMF. This highlights the importance of including a model for the relationship between subhalo and satellite galaxy size like the one used in our analysis. As discussed in Section \ref{sec:mw}, we also marginalize over MW halo mass and the efficiency of subhalo disruption due to baryonic physics, which are both key systematics in the MW satellite inference. Our MW host halo mass marginalization procedure is analytic, unlike the simulation-based method employed in \cite{Newton201108865}, due to the limited statistics of MW-like simulations that include realistic LMC analogs. The significant improvements in sensitivity to non-CDM physics afforded by modeling the LMC satellite system further reinforce the importance of simulation suites of MW-like systems including realistic LMC analog systems.

On the strong lensing side, the \cite{Gilman190806983} study upon which we base our analysis uses flux ratio statistics that are significantly more constraining than the gravitational imaging data underlying the \cite{Enzi201013802} joint analysis. This additional constraining power results from the fact that current gravitational imaging data probes $\sim 10^9\ M_{\mathrm{\odot}}$ subhalos while flux ratio anomalies are sensitive to the presence of lower-mass subhalos. In terms of modeling, \cite{Gilman190806983} explicitly account for the host mass and redshift dependence of the SHMF using \texttt{Galacticus}---these are leading-order effects in predicting the SHMF for a given lens and its lens-to-lens variation---while the \cite{Vegetti180101505} and \cite{Riitondale181103627} analyses that the joint constraints in \cite{Enzi201013802} are based on do not. In addition, \cite{Gilman190806983} self-consistently account for the reduction in halo concentration in WDM, which significantly increases the sensitivity of lensing observations to WDM effects and also models the effects of tidal stripping on subhalos after infall, which is again crucial to accurately forward-model flux ratio observations.

Finally, we emphasize the following key aspects of our probe combination relative to the procedure in \cite{Enzi201013802}, which combines fully marginalized one-dimensional $M_{\mathrm{hm}}$ distributions from various small-scale structure probes including MW satellites and gravitational imaging to derive joint WDM constraints:
\begin{enumerate}
\item We cast the subhalo populations inferred from MW satellites and from the group-mass, $z\sim 0.5$ host halos probed by strong lensing into a common, multidimensional parameter space of projected subhalo number density $\Sigma_{\mathrm{sub}}$ versus WDM half-mode mass $M_{\mathrm{hm}}$;
\item We combine these $\Sigma_{\mathrm{sub}}$--$M_{\mathrm{hm}}$ distributions to construct a joint marginal likelihood that is strictly more constraining and informative than the joint $M_{\mathrm{hm}}$ distribution resulting from fully marginalizing over all additional parameters (see the right panel of Figure \ref{fig:1d_hists_systematics}), improving the precision of our joint analysis; and
\item We model the differential efficiency of subhalo disruption due to the central galaxies in the different host halo mass and redshift regimes probed by MW satellites and strong lensing, finding that our results are robust to uncertainties in these effects at the $\sim 10\%$ level, which lends confidence to the robustness of our results.
\end{enumerate}

The differences in the underlying data used in our inference---and particularly the inclusion of LMC-associated satellites in the MW satellite analysis and the use of strong lensing flux ratio statistics that probe lower-mass subhalos than current gravitational imaging data---therefore result in more precise joint constraints than those obtained in \cite{Enzi201013802} and allow us to significantly improve upon their WDM limit. Moreover, the joint analysis choices described above lend to the robustness and accuracy of our results.


\section{Conclusions}
\label{sec:conclusion}

In this paper, we performed a multidimensional joint analysis of the distribution of small-scale dark matter structure inferred from MW satellite galaxies and strong gravitational lensing. In particular, we combined state-of-the-art dark matter substructure measurements derived from (i) the MW satellite galaxy population over $\sim 75\%$ of the sky and (ii) the flux ratio statistics and image positions from eight quadruply imaged quasars. By combining constraints on the projected subhalo number density and the half-mode mass describing the suppression of the subhalo mass function in thermal relic WDM, we improved lower limits on the WDM particle mass derived independently by breaking degeneracies among the inferred subhalo distributions at each scale for the first time. Our $m_{\mathrm{WDM}}$ constraint is more stringent than any limit set by independent analyses of small-scale structure probes to date.

Our key results are summarized below:
\begin{enumerate}
\item Our multidimensional joint analysis extracts information that was not accessed by MW satellite or strong lensing analyses independently, improving WDM constraints by \textcolor{black}{$\sim 30\%$}, with \textcolor{black}{$M_{\mathrm{hm}}<10^{7.0}\ M_{\mathrm{\odot}}$} (\textcolor{black}{$m_{\mathrm{WDM}}>9.7\ \mathrm{keV}$}) at $95\%$ confidence, or \textcolor{black}{$M_{\mathrm{hm}}=10^{7.4}\ M_{\mathrm{\odot}}$} (\textcolor{black}{$m_{\mathrm{WDM}}=7.4\ \mathrm{keV}$}) disfavored with a 20:1 marginal likelihood ratio. (Figures \ref{fig:combined_hist}--\ref{fig:1d_hists});

\item Our joint WDM constraint is robust to uncertainties in the differential efficiency of subhalo disruption between the MW and strong lens host halo mass and redshift regimes at the $\sim 10\%$ level;

\item Projected subhalo number density constraints from MW satellites and strong lensing flux ratio statistics are in mild tension but are sensitive to uncertainties in the efficiency of subhalo disruption in the corresponding host halo mass and redshift regimes;

\item We discuss key systematics that are conservatively marginalized over in the current analysis but which must be mitigated in future work to claim a detection of non-CDM physics from small-scale structure measurements. These systematics include the line-of-sight contribution to the strong lensing signal, the differential efficiency of subhalo disruption due to baryons at the MW and lensing host halo mass and redshift scales, and the properties of the MW and strong lens host halos (Figure \ref{fig:1d_hists_systematics});

\item Inferences of the small-scale dark matter structure from MW satellites and strong lensing are consistent despite the completely different nature of these probes and differences in their corresponding host halo mass and redshift regimes.

\end{enumerate}

Recent studies have identified a variety of microphysical dark matter properties that suppress small-scale structure in a manner quantitatively similar to WDM, including the strength of velocity-independent interactions between dark matter and protons \citep{Nadler190410000}, the production mechanism of nonthermal dark matter in early matter-dominated cosmologies \citep{Miller190810369}, and the dark matter formation redshift in models of ``late-forming'' dark matter \citep{Das201001137}. Our jointly derived WDM constraints directly inform all of these properties. Dark matter models that feature qualitatively different suppression of small-scale structure compared to WDM can also be constrained by constructing a conservative mapping; for example, such mappings have been applied to constrain fuzzy dark matter \citep{Schutz200105503}, models with velocity-dependent self- and Standard Model dark matter interactions \citep{Tulin170502358,Maamari201002936}, and models within the ETHOS framework \citep{Bohr200601842}. Such dark matter physics may manifest differently in small-scale structure probes like MW satellites and strong lensing that are sensitive to halo abundances and concentrations in unique ways, and we regard this as a particularly compelling avenue for future work.

We expect the relative improvement offered by our probe combination to continue to increase as both techniques progress due to both additional data from existing instruments and next-generation observational facilities. Excitingly, the sample sizes of both nearby ultrafaint dwarf galaxies and quadruply lensed quasars are expected to drastically increase with LSST \citep{Ivezic08052366}, Euclid Space Telescope \citep{Euclid}, and Nancy Grace Roman Space Telescope \citep{Roman} observations. Forthcoming facilities including the Maunakea Spectroscopic Explorer \citep{MSE:2019} will also help to confirm the nature of candidate MW satellites and faint dwarf galaxies throughout the Local Volume, while wide-aperture and extremely large telescopes (ELTs) will provide detailed information about the dynamical masses of these systems, which is key to refine galaxy--halo connection and WDM constraints \citep{Simon190304743}. Meanwhile, the unprecedented sample of strong lenses expected to be discovered within 5--10 yr will yield precise measurements of the differential line-of-sight and substructure contributions to lensing signal and will allow the selection functions of strong lenses to be better quantified. Observations of extended source emission will also help constrain lens macromodels.

With sufficiently stringent limits on the minimum luminous halo mass from nearby dwarf galaxies and on the mass scale of a cutoff in the subhalo mass function, our procedure for combining satellite galaxy and strong lensing posteriors can potentially provide evidence for the existence of dark subhalos---i.e., subhalos devoid of observable baryonic components---which are a key, unverified prediction of many viable dark matter models. \cite{Nadler191203303} estimate that the lowest-mass halo expected to host a dwarf galaxy is more massive than $\sim 10^7\ M_{\mathrm{\odot}}$. Thus, the future observations discussed above, which are expected to constrain the subhalo mass function at and below these mass scales, will either yield evidence for a cutoff in galaxy (or halo) formation or evidence for halos devoid of observable baryonic matter. We plan to pursue these measurements by developing a multipronged theoretical framework to jointly infer the distribution of small-scale structure using heterogeneous data.


\acknowledgments

We are grateful to Yao-Yuan Mao for sharing the \cite{Mao150302637} Milky Way zoom-in simulation suite data. We thank Sebastian Wagner-Carena for useful conversations and Keith Bechtol, Alex Drlica-Wagner, and Yao-Yuan Mao for comments on the manuscript. 

This research received support from the National Science Foundation (NSF) under grant No.\ AST-1716527 to R.H.W.\ and grant No.\ NSF DGE-1656518 through the NSF Graduate Research Fellowship received by E.O.N. A.J.B.\ and X.D.\ acknowledge support from NASA ATP grant 17-ATP17-0120. T.T.\ acknowledges support by the Packard Foundation through a Packard Research Fellowship, by the NSF through grants AST-1836016 and AST-1714953, and by the Gordon and Betty Moore Foundation. This research used \url{https://arXiv.org} and NASA's Astrophysics Data System for bibliographic information.

\bibliographystyle{yahapj2}
\bibliography{references}


\begin{figure*}[t!]
    \hspace{-1mm}
    \includegraphics[scale=0.43]{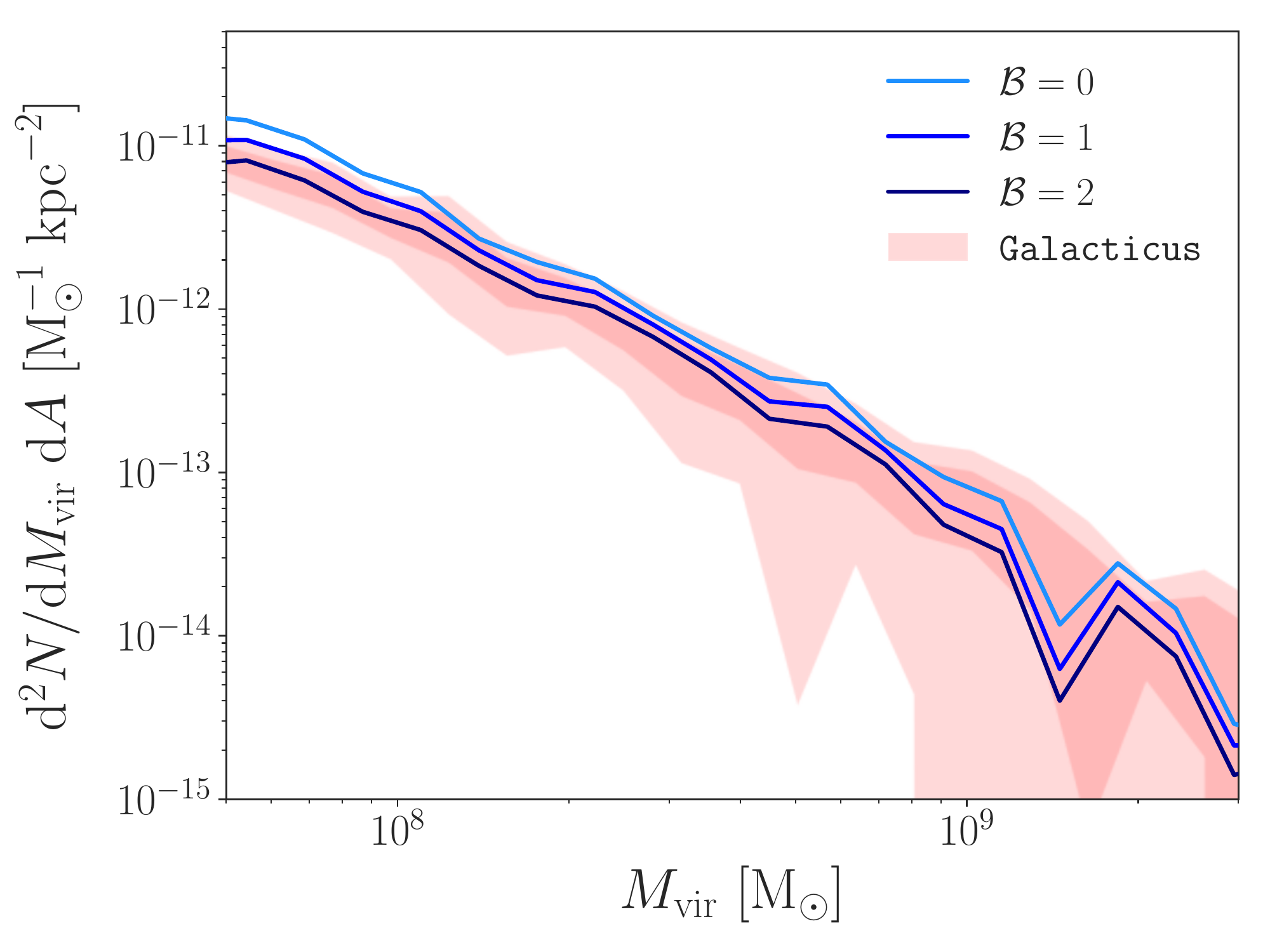}
    \hspace{2mm}
    \includegraphics[scale=0.43]{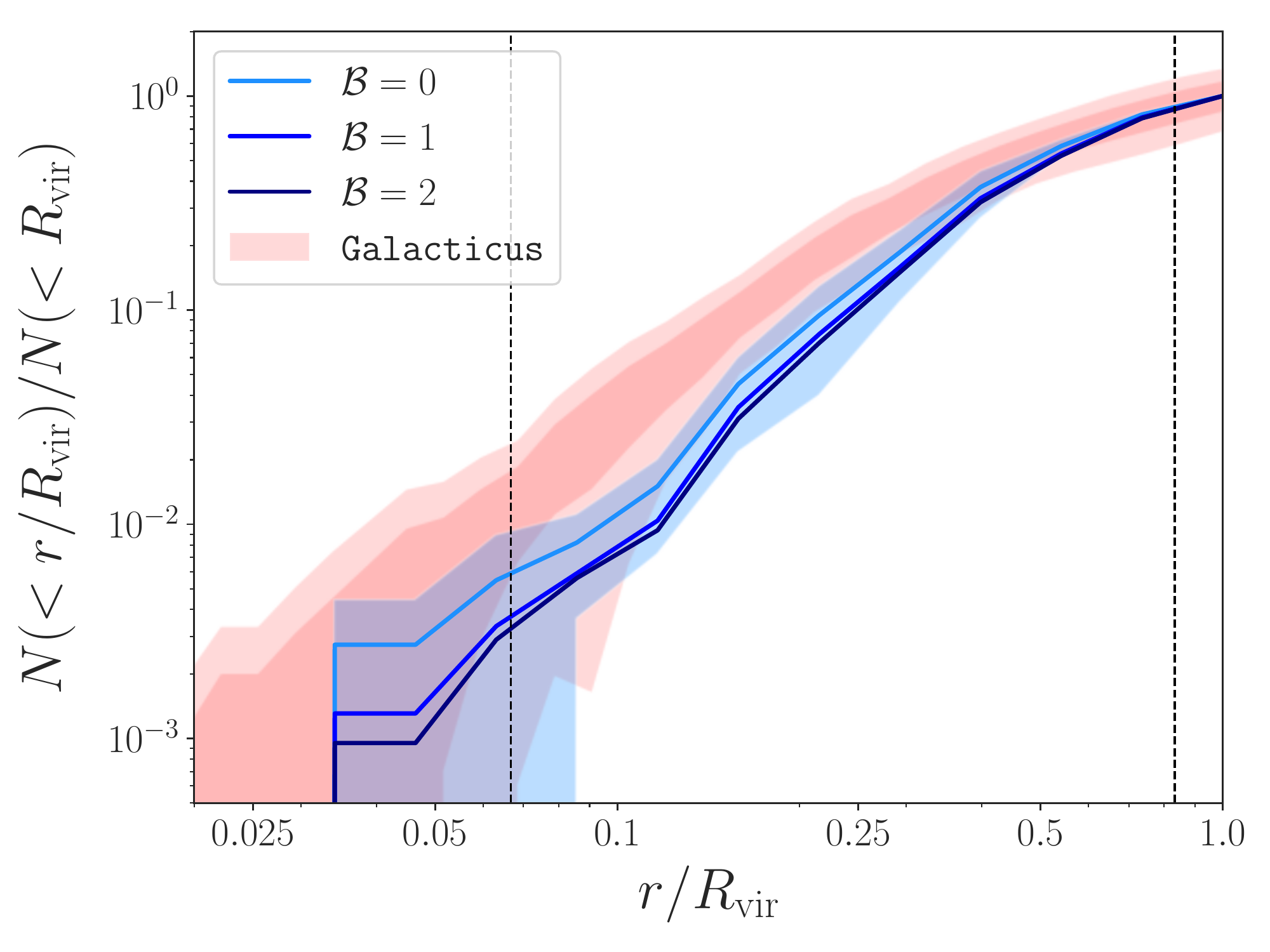}
    \caption{Left panel: projected subhalo mass function for MW-like host halos as a function of present-day subhalo virial mass. Blue lines show results from the zoom-in simulations used in our MW satellite inference for several values of the baryonic disruption efficiency parameter $\mathcal{B}$ (models with more efficient subhalo disruption are shown in darker colors). Right panel: average radial subhalo distribution in units of the host halo virial radius for our MW-like simulations (blue). Dashed vertical lines approximately mark the radial range of observed MW satellite galaxies used in our analysis. In both panels, dark (light) red contours show $68\%$ ($95\%$) confidence intervals from \texttt{Galacticus} for a sample of halos with characteristics matched to our MW-like simulations (see Appendix \ref{sec:comparison} for details). To calculate the \texttt{Galacticus} radial distributions, we only consider halos with $M_{\mathrm{peak}}>10^8\ M_{\mathrm{\odot}}$ in addition to the $V_{\mathrm{peak}}$ and $V_{\mathrm{max}}$ cuts described in Appendix \ref{sec:sim_details} to facilitate a direct comparison to our simulation results.}
    \label{fig:comparison_mvir}
\end{figure*}

\appendix

\section{Milky Way Zoom-in Simulations}
\label{sec:sim_details}

Our realistic MW-like simulations are drawn from the suite of 45 zoom-in simulations presented in \cite{Mao150302637}, which have host halo virial masses between $1.2$ and $1.6\times 10^{12}\ M_{\rm{\odot}}$.\footnote{We define virial quantities according to the \cite{Bryan_1998} virial definition, with overdensity $\Delta_{\rm vir}\simeq 99.2$ in units of the critical density as appropriate for our fiducial cosmological parameters.} The highest-resolution particles in these simulations have a mass of $3\times 10^{5}\ M_{\rm \odot}\ h^{-1}$, and the softening length in the highest-resolution regions is $170\ \mathrm{pc}\ h^{-1}$. Subhalos in these simulations are well resolved down to a present-day maximum circular velocity of $V_{\rm max} \approx 9\ \rm{km\ s}^{-1}$ \citep{Mao150302637}, and halo catalogs and merger trees were generated using the {\sc Rockstar} halo finder and the {\sc consistent-trees} merger code \citep{Behroozi11104372,Behroozi11104370}.

To account for the limited resolution of these simulations, we only analyze subhalos with maximum circular velocity $V_{\mathrm{max}}>9\ \mathrm{km\ s}^{-1}$ and peak maximum circular velocity $V_{\rm peak} > 10\ \rm{km\ s}^{-1}$, which are typically resolved with $\gtrsim 100$ particles at the time $V_{\mathrm{peak}}$ is achieved. In addition, because we construct the $\mathcal{B}$--$\Sigma_{\mathrm{sub}}$ relation by conservatively matching the MW zoom-in and \texttt{Galacticus}-predicted SHMFs down to the minimum halo mass scale, which corresponds to $M_{\mathrm{peak}}>2.5\times 10^8\ M_{\mathrm{\odot}}$ or $V_{\mathrm{peak}}>19\ \mathrm{km\ s}^{-1}$ before accounting for MW host halo mass uncertainty \citep{Nadler191203303}, only subhalos resolved with greater than $\sim 600$ particles at the time $V_{\mathrm{peak}}$ is achieved directly influence our results.

\begin{figure*}[t!]
    \hspace{-1mm}
    \includegraphics[scale=0.43]{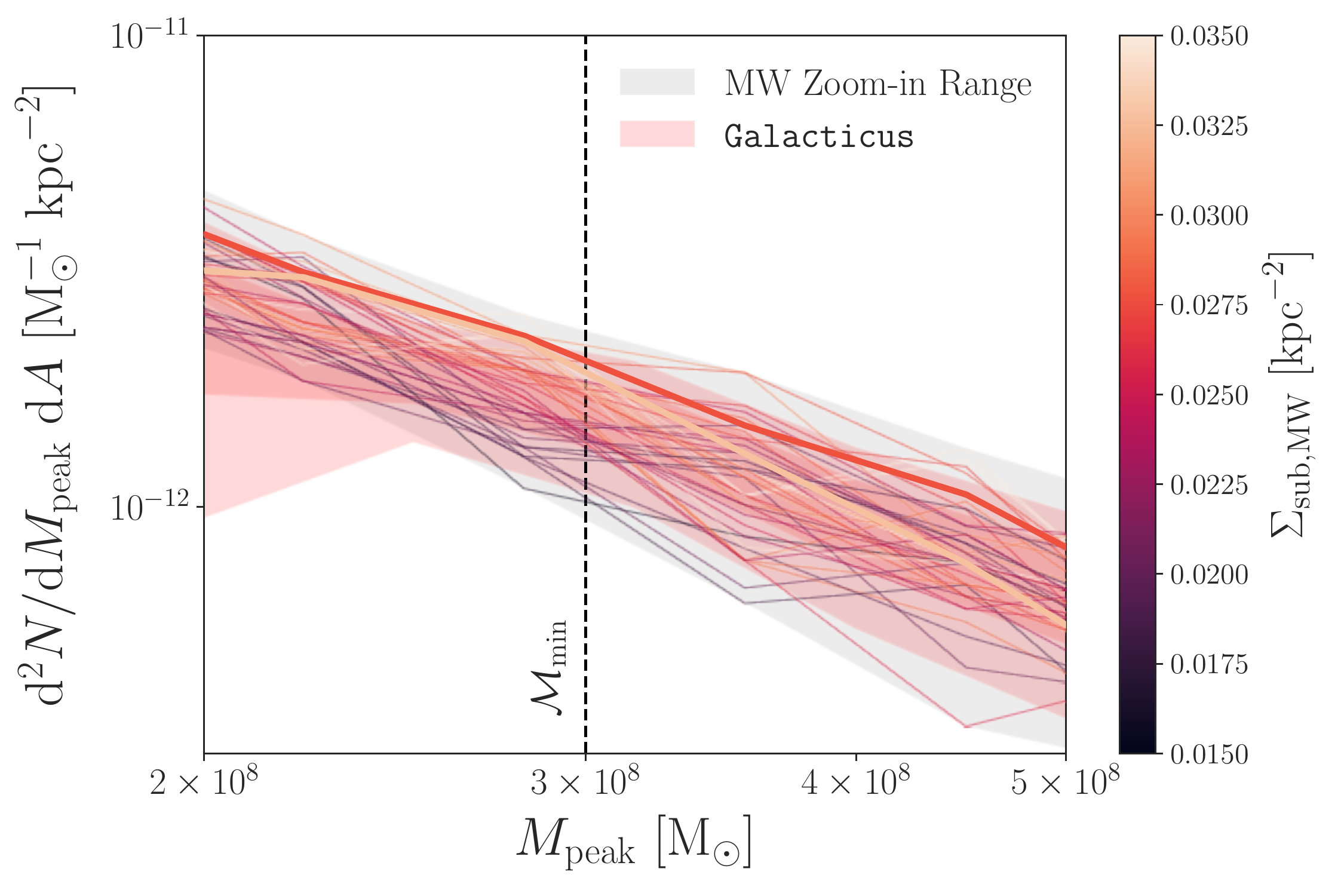}
    \hspace{2mm}
    \includegraphics[scale=0.43]{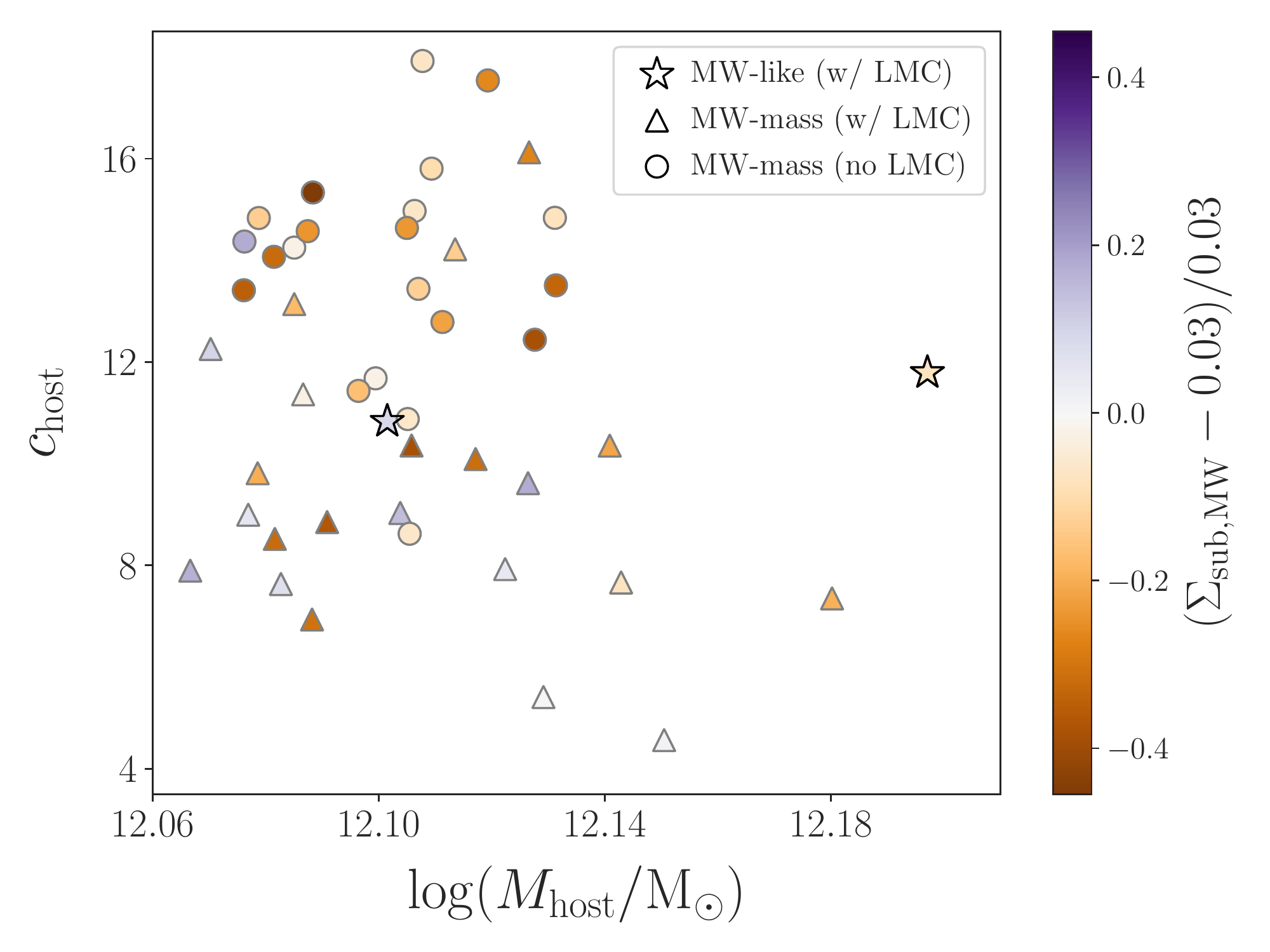}
    \caption{Left panel: projected subhalo mass functions versus peak subhalo virial mass. Lines correspond to individual zoom-in simulations from the \cite{Mao150302637} suite of MW-mass host halos and are colored according to their projected subhalo number density $\Sigma_{\mathrm{sub,MW}}$; the two thickest lines correspond to the MW-like simulations used in our analysis. The gray band indicates the range of SHMFs from these simulations, and dark (light) red contours show $68\%$ ($95\%$) confidence intervals from \texttt{Galacticus} for a sample of halos with characteristics matched to our MW-like simulations. Right panel: relation between host halo mass, concentration, and $\Sigma_{\mathrm{sub,MW}}$ for the same suite of zoom-in simulations. Stars show the two MW-like simulations used in our analysis, which include realistic LMC analog systems, triangles show simulations from this suite that have an LMC analog (i.e., a subhalo with $V_{\mathrm{max}}>55\ \mathrm{km\ s}^{-1}$) anywhere within their virial radius, and circles show simulations that do not have an LMC analog. Colors indicate fractional differences relative to the average value of $\Sigma_{\mathrm{sub,MW}}$ from the two MW-like simulations.}
    \label{fig:host_scatter}
\end{figure*}

As noted in Section \ref{sec:bridge}, we include disrupted orphan subhalos in our predictions using the model presented in \cite{Nadler180905542}. This model semianalytically tracks the orbital evolution of subhalos after disruption while accounting for tidal stripping and the evolving potential of the host halo, and it is calibrated by comparing to higher-resolution versions of halos from the \cite{Mao150302637} zoom-in simulation suite. We used a higher-resolution resimulation of one of our realistic MW-like halos (described in \citealt{Nadler191203303}), which is run with high-resolution particles of $4\times 10^4\ M_{\mathrm{\odot}}$ and an $85\ \mathrm{pc}\ h^{-1}$ minimum softening length, to check that our SHMF predictions and the resulting $\mathcal{B}$--$\Sigma_{\mathrm{sub}}$ relation are numerically converged when including orphans.

As demonstrated in \cite{Nadler180905542}, orphans contribute to the subhalo population at our fiducial zoom-in resolution at the $\sim 10\%$ level. The orphan contribution is roughly mass independent, increases at small Galactocentric radii, and is not highly degenerate with WDM physics, which suppresses low-mass subhalos in a radially independent manner. Furthermore, \cite{Nadler191203303} show that the addition of orphans does not significantly affect galaxy--halo connection constraints derived from DES and PS1 data. The development of a self-consistent orphan model that can be applied to both $N$-body simulations and \texttt{Galacticus} predictions is left to future work.

\section{Comparing Milky Way Zoom-in Simulations to \texttt{Galacticus}}
\label{sec:comparison}

We construct \texttt{Galacticus} predictions corresponding to our realistic MW-like simulations by generating host halos from a mass range corresponding to the simulations described in Appendix \ref{sec:sim_details}. Host halo concentrations are generated using the \cite{Diemer180907326} mass--concentration relation with $0.16\ \mathrm{dex}$ scatter and span the concentration values of the hosts in our MW-like zoom-in simulations. Out of these runs, we select halos that satisfy:
\begin{enumerate}
\item Host halo NFW concentration of $7 < c_{\mathrm{host}} < 16$;
\item A realistic LMC analog system that accretes within the last $2\ \mathrm{Gyr}$, has a present-day maximum circular velocity of $V_{\mathrm{max}} > 55 \mathrm{km\ s}^{-1}$, Galactocentric distance of $40\ \mathrm{kpc} < D < 60\ \mathrm{kpc}$, and Galactocentric velocity of $267\ \mathrm{km\ s}^{-1} < V < 375\ \mathrm{km\ s}^{-1}$;
\item A Gaia-Enceladus-like accretion event, i.e., a merger with a satellite-to-host mass ratio in the range
$[0.15,0.25]$ in the redshift range $1<z<2$.
\end{enumerate}

These criteria are chosen to match those imposed on our realistic MW-like simulations \citep{Nadler191203303}. Note that we used Galactocentric distance to define LMC properties rather than heliocentric distance as in \cite{Nadler191203303}, but we do not expect this choice to impact our results. With the above criteria, roughly $0.1\%$ of \texttt{Galacticus} runs in the relevant host halo mass range are accepted and we are left with 14 independent realizations. For computational efficiency, we ignore all subhalos accreted earlier than $z=5$ when generating \texttt{Galacticus} predictions. A negligible fraction of halos that accrete earlier than $z=5$ survive in our $N$-body simulations, implying that this choice does not impact our comparisons. Furthermore, we self-consistently employ the subhalo $V_{\mathrm{peak}}$ and $V_{\mathrm{max}}$ cuts described in Appendix \ref{sec:sim_details} when comparing to our simulation results.

\begin{figure*}[t!]
    \hspace{-10mm}
    \includegraphics[scale=0.48]{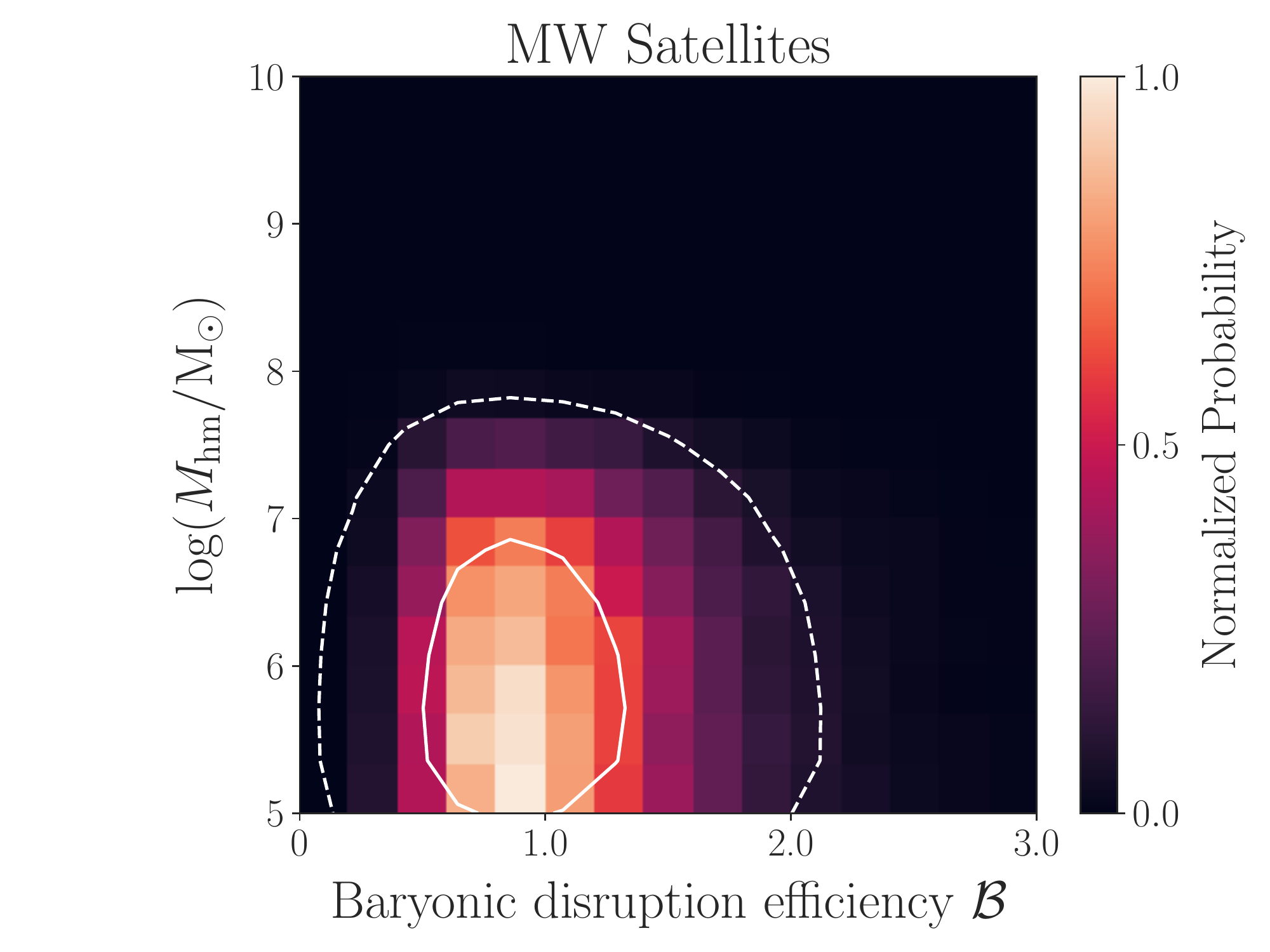}
    \includegraphics[scale=0.48]{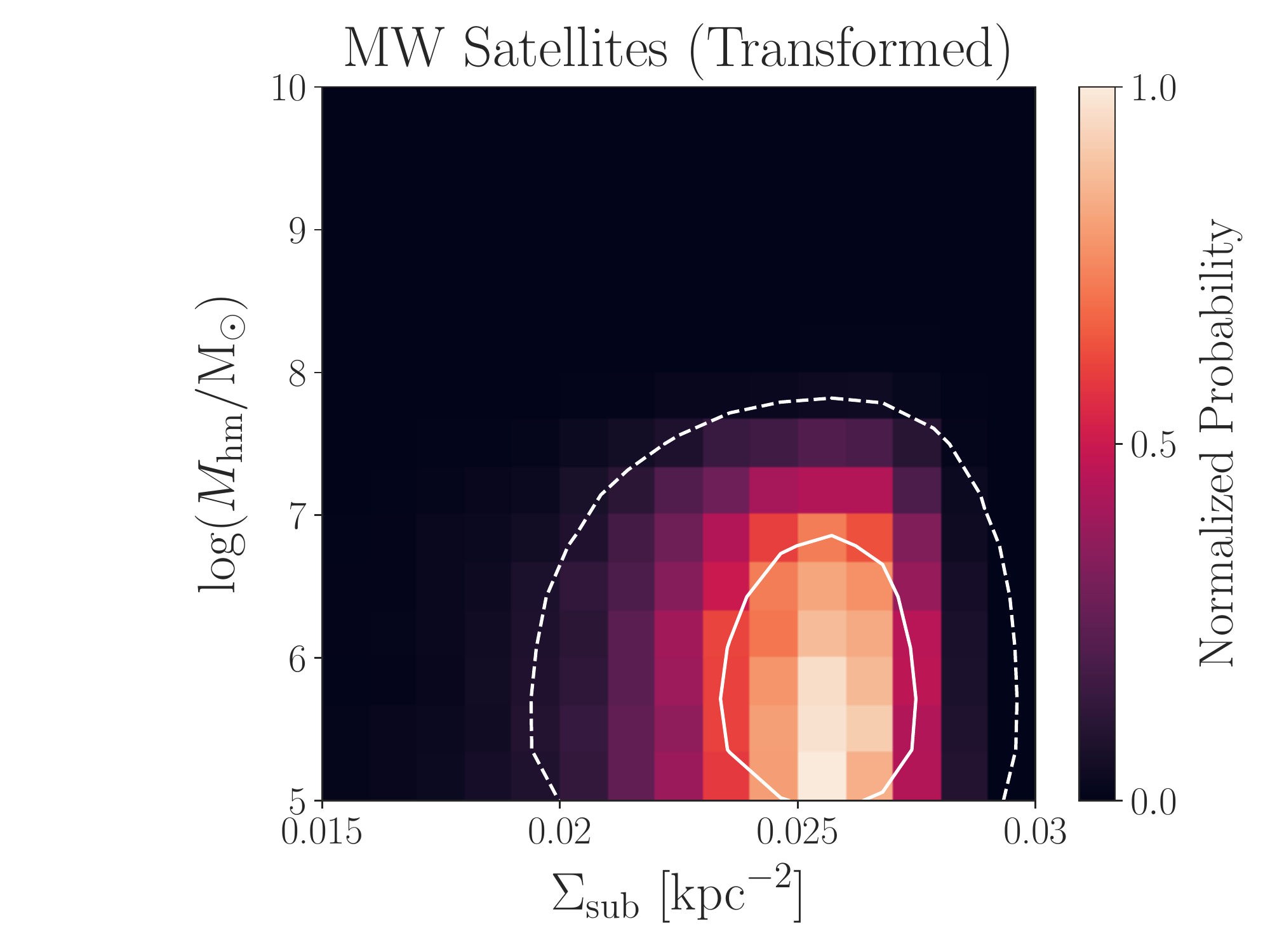}
    \caption{Left panel: posterior distribution of WDM half-mode mass versus baryonic disruption efficiency from our analysis of the MW satellite population using a lognormal prior on $\mathcal{B}$, rather than a uniform prior as in Figure \ref{fig:satellite_hist_raw}. Right panel: corresponding posterior distribution of WDM half-mode mass versus projected subhalo number density at the strong lensing scale, inferred according to the procedure in Section \ref{sec:bridge} with $q=1$ (see the left panel of Figure \ref{fig:transformed_hist} for comparison). In both panels, color maps show the probability density normalized to its maximum value in each parameter space, and solid (dashed) white lines indicate $1\sigma$ ($2\sigma$) contours for a two-dimensional Gaussian distribution.}
    \label{fig:prior_hist}
\end{figure*}

The left panel of Figure \ref{fig:comparison_mvir} shows that, in addition to the agreement among the SHMFs as a function of $M_{\mathrm{peak}}$ demonstrated in Figure \ref{fig:shmf}, \texttt{Galacticus} predictions are consistent with our zoom-in simulations for SHMFs evaluated using present-day subhalo virial mass. This indicates that the amount of stripping experienced by subhalos in our $N$-body simulations is well captured by the \texttt{Galacticus} model, on average. However, as shown in the right panel of Figure \ref{fig:comparison_mvir}, our zoom-in simulations yield radial subhalo distributions that are slightly less concentrated than those predicted by \texttt{Galacticus}. This discrepancy is unchanged when comparing to the higher-resolution version of one of our simulations described above. We note that increasing the radial concentration of the subhalo distribution predicted by our simulations at fixed subhalo abundance would further strengthen our minimum halo mass and WDM constraints \citep{Nadler191203303}. Because the radial distribution in our simulations and \texttt{Galacticus} are respectively subject to subtle numerical uncertainties including artificial subhalo disruption and semianalytic modeling of dynamical friction,  we plan to explore this discrepancy systematically in future work.

\section{Host-to-host, Poisson, and projection scatter in $\Sigma_{\mathrm{sub,MW}}$}
\label{sec:scatter}

As we have emphasized, it is challenging to accurately estimate the host-to-host scatter in $\Sigma_{\mathrm{sub,MW}}$ given the requirements we place on our realistic MW-like simulations. In this Appendix, we take a very conservative approach by quantifying the scatter in $\Sigma_{\mathrm{sub,MW}}$ for the entire suite of simulations from \cite{Mao150302637} described in Appendix \ref{sec:sim_details}. In particular, the left panel of Figure \ref{fig:host_scatter} shows the projected SHMF for all 45 of the \cite{Mao150302637} simulations, shaded by their $\Sigma_{\mathrm{sub,MW}}$ calculated according to the procedure in Section \ref{sec:b--sigma_sub}, and the right panel of Figure \ref{fig:host_scatter} shows the dependence of $\Sigma_{\mathrm{sub,MW}}$ on host halo mass and concentration. These panels illustrate that the scatter in $\Sigma_{\mathrm{sub,MW}}$ is at most $\sim 40\%$ toward smaller values of $\Sigma_{\mathrm{sub,MW}}$ than inferred from our realistic MW-like simulations, and at most $\sim 20\%$ toward larger values of $\Sigma_{\mathrm{sub,MW}}$. The dependence of subhalo abundance on host halo properties among these zoom-in simulations is studied in detail by \cite{Mao150302637} and \cite{Fielder180705180}.

Although the host-to-host uncertainty quoted above is not small in an absolute sense, the scatter in either direction is overshadowed by the factor of 2 uncertainty introduced by $q$ in the translation from $\Sigma_{\mathrm{sub,MW}}$ to $\Sigma_{\mathrm{sub}}$. Moreover, scatter toward lower values of $\Sigma_{\mathrm{sub,MW}}$ (which is more common) would further strengthen our joint WDM constraints as described in Section \ref{sec:vary_q}. Furthermore, this estimate of the host-to-host scatter using the entire zoom-in simulation suite is an overestimate because it does not leverage additional information about the properties of the MW halo and because the \cite{Mao150302637} hosts were chosen to span a cosmologically representative range of formation histories rather than being selected uniformly in host halo mass. We therefore regard our current analysis to be conservative because it accounts for the dominant uncertainties (i.e., $M_{\mathrm{MW}}$ and $q$), and we plan to simultaneously infer $q$, $M_{\mathrm{MW}}$, and $\Sigma_{\mathrm{sub,MW}}$ along with their associated uncertainties in future work.

The Poisson scatter in the projected SHMFs predicted by our simulations near the minimum halo mass is also moderate compared to the other systematic uncertainties discussed above. In particular, given our fiducial binning scheme, there are $\sim 50$ subhalos per $M_{\mathrm{peak}}$ bin near $\mathcal{M}_{\mathrm{min}}=3\times 10^8\ M_{\mathrm{\odot}}$, corresponding to $\sim 15\%$ Poisson scatter, which is again relatively minor compared to uncertainties in $q$ and $M_{\mathrm{MW}}$. We refer the reader to \cite{Mao150302637} for a detailed study of these subhalo populations that justifies the use of a Poisson distribution to describe their scatter.

Finally, we note that the scatter in the projected subhalo mass function induced by different orientations for the projection of the MW subhalo population is also small compared to the other sources of uncertainty we have discussed. For example, the subhalo population projected with half of the virial radius in our MW-like simulations varies at the percent level for different orientations.

\section{Subhalo Disruption Efficiency Prior}
\label{sec:priors}

To formulate our probe combination in a statistically consistent way, we reran the \cite{Nadler200800022} MW satellite analysis with a uniform prior on the baryonic subhalo disruption $\mathcal{B}$ as described in Section \ref{sec:mw}. However, the fiducial \cite{Nadler200800022} model assumes a lognormal prior on this quantity centered around $\mathcal{B}=1$ (i.e., the expectation for the efficiency of subhalo disruption from hydrodynamic simulations of MW-mass halos; also see \citealt{Nadler180905542,Nadler191203303}). In this appendix, we explore the effects of performing the probe combination using this lognormal prior.

In particular, Figure \ref{fig:prior_hist} shows the posterior from the MW satellite analysis in the $\mathcal{B}$--$M_{\mathrm{hm}}$ and $\Sigma_{\mathrm{sub}}$--$M_{\mathrm{hm}}$ parameter spaces, translated according to Equation \ref{eq:q} with $q=1$, assuming the fiducial \cite{Nadler200800022} prior of $\ln \mathcal{B}\sim \mathcal{N}(\mu=1,\sigma=0.5)$. It is visually evident that this prior favors a narrower range of $\Sigma_{\mathrm{sub}}$, as expected. Using this alternative prior and setting $q=1$ does not change the results of our joint analysis, with \textcolor{black}{$M_{\mathrm{hm}}<10^{7.0}\ M_{\mathrm{\odot}}$} (\textcolor{black}{$m_{\mathrm{WDM}}>9.7\ \mathrm{keV}$}) at $95\%$ confidence and \textcolor{black}{$M_{\mathrm{hm}}=10^{7.5}\ M_{\mathrm{\odot}}$} (\textcolor{black}{$m_{\mathrm{WDM}}=6.9\ \mathrm{keV}$}) disfavored with a 20:1 marginal likelihood ratio. This is due to a cancellation of effects: using a lognormal prior on $\mathcal{B}$ slightly strengthens our MW satellite constraint on $M_{\mathrm{hm}}$ (compare the left panel of Figure \ref{fig:prior_hist} to Figure \ref{fig:satellite_hist_raw}), but also removes the low-$\Sigma_{\mathrm{sub}}$ tail of the MW satellite posterior (compare the right panel of Figure \ref{fig:prior_hist} to the left panel of Figure \ref{fig:combined_hist}). Because larger values of $\Sigma_{\mathrm{sub}}$ lead to weaker joint constraints as described in Section \ref{sec:vary_q}, these effects push our joint WDM constraints in opposite directions and happen to be roughly equal in magnitude.

\end{document}